\documentclass[11pt]{article}
\pdfoutput=1
\usepackage{jpub,amsmath,amssymb,hyperref}
\usepackage{aas_macros}
\usepackage{bm}
\usepackage{placeins}
\usepackage{bbold}
\usepackage{pgfplots,todonotes}
\usepackage{tikz}
\usepackage[latin1]{inputenc}
\usepackage[normalem]{ulem}
\usetikzlibrary{arrows,positioning,decorations.markings,decorations.pathmorphing,calc}
\definecolor{hyperref}{RGB}{026,028,087}
\def\gsim{ \lower .75ex \hbox{$\sim$} \llap{\raise .27ex \hbox{$>$}} }
\def\lsim{ \lower .75ex \hbox{$\sim$} \llap{\raise .27ex \hbox{$<$}} }
\def\be{\begin{equation}}
\def\ee{\end{equation}}
\def\bea{\begin{eqnarray}}
\def\eea{\end{eqnarray}}

\newcommand{\commentout}[1]{}

\newcommand{\MPl}{M_{\rm Pl}}

\newcommand{\comment}[1]{}

\newcommand{\bs}{\begin{split}}

\newcommand{\nn}{\nonumber}

\newcommand{\blue}[1]{\textcolor{black}{#1}}

\def\({\left(}
\def\){\right)}

\newcommand*{\mathcolor}{}
\def\mathcolor#1#{\mathcoloraux{#1}}
\newcommand*{\mathcoloraux}[3]{%
\protect\leavevmode
\begingroup
\color#1{#2}#3%
\endgroup
}
\newlength{\stheight}
\newcommand\textst[1][fu-grey]{
\ifmmode\setlength{\stheight}{+1.0ex}
\else\setlength{\stheight}{+0.5ex}
\fi
\bgroup\markoverwith{\textcolor{#1}{\rule[\the\stheight]{2pt}{1.0pt}}}\ULon
} 
\newcommand{\textins}[2][fu-grey]{
\ifmmode\mathcolor{#1}{#2}
\else\textcolor{#1}{#2}\@\,
\fi
}
\graphicspath{{./}}
\allowdisplaybreaks

\usepackage{titlesec}
\titleformat{\part}{\Large\bfseries}{}{0pt}{Part \thepart\ --\ }

\tikzstyle{vecArrow} = [thick, decoration={markings,mark=at position
1 with {\arrow[semithick]{open triangle 60}}},
double distance=1.4pt, shorten >= 5.5pt,
preaction = {decorate},
postaction = {draw,line width=1.4pt, white,shorten >= 4.5pt}]

\begin{document}
\title{Positivity bounds from multiple vacua and their cosmological consequences}

\author[a]{Scott Melville,}
\author[b,a]{Johannes Noller}

\affiliation[a]{DAMTP, University of Cambridge, Wilberforce Road, Cambridge, CB3 0WA, U.K.}
\affiliation[b]{Institute of Cosmology \& Gravitation, University of Portsmouth, Portsmouth, PO1 3FX, U.K.}

\emailAdd{scott.melville@damtp.cam.ac.uk}
\emailAdd{johannes.noller@port.ac.uk}

\date{today}
\abstract{
Positivity bounds---constraints on any low-energy effective field theory imposed by the fundamental axioms of unitarity, causality and locality in the UV---have recently been used to constrain various effective field theories relevant for cosmology.
However, to date most of these bounds have assumed that there is a single Lorentz-invariant vacuum in which all fields have zero expectation value and in many cosmologically relevant models this is not the case. 
We explore ways to overcome this limitation by investigating a simple example model, the covariant Galileon, which possesses a one-parameter family of Lorentz-invariant vacua as well as multiple boost-breaking vacua. 
Each of these vacua has a corresponding set of positivity bounds, and we show how a particular (beyond-the-forward-limit) bound can be used to map out the parameter space according to which vacua may persist in the UV theory, finding that in general there are regions in which none, one or many of the effective field theory vacua can be consistent with unitarity, causality and locality in the UV. 
Finally, we discuss the interplay between this map and cosmological observations. 
We find that the observationally favoured region of parameter space is incompatible with a large class of vacua, and conversely that particular boost-breaking vacua would imply positivity bounds that rule out otherwise observationally favoured cosmologies.
We also identify a specific boost-breaking vacuum which is ``closest'' to the cosmological background, and show that the particular positivity bound we consider reduces the otherwise cosmologically favoured region of Galileon parameter space by up to $70 \%$, ruling out the vast majority of cosmologies with a positive coefficient for the cubic Galileon in the process. 
}

\keywords{Positivity Bounds, Effective Field Theory, Dark Energy}

\setcounter{tocdepth}{2}
\maketitle

\section{Introduction}

General Relativity (GR) is a cornerstone of $20^{\rm th}$ century physics, having proven remarkably successful over a wide range of scales. 
However, since it is ultimately an effective field theory (EFT)---breaking down at the Planck scale if not before---and can encounter difficulties when accounting for the observed late-time acceleration \cite{Riess:1998cb, Perlmutter:1998np}---the well-known cosmological constant problem \cite{Weinberg:1988cp}---we know that it cannot be the fundamental description of our Universe.
This need to explore possible deviations from GR has led to a plethora of different models which introduce additional dark sector fields. 

~

The simplest addition beyond the tensor degrees of freedom of GR is a single scalar degree of freedom, $\phi$, which can play the role of dark energy in the late Universe (see \cite{Capozziello:2007ec, Capozziello:2011et, Clifton:2011jh, Joyce:2014kja, Bull:2015stt, Koyama:2015vza} for reviews). 
The next generation of experiments will improve the precision of our cosmological observations and provide increasingly tight constraints on the parameters in these models.
But without further theoretical guidance, it is not clear which models are the best motivated or should guide future survey strategies, or how our observations of these EFT parameters should shape future UV model-building. 

~

``Positivity bounds'' can provide this much-needed sign-posting, mapping out the landscape of consistent low-energy EFTs. These are constraints which must be satisfied by the EFT scattering amplitudes as a consequence of unitarity, causality and locality---the foundational axioms of quantum field theory---in the UV \cite{Adams:2006sv, Nicolis:2009qm, deRham:2017avq}\footnote{
In this work we focus on linear positivity bounds for a single scalar field. 
There has been much progress recently developing similar (and in some cases stronger) bounds for spinning particles \cite{Bellazzini:2016xrt, deRham:2017zjm, Remmen:2020uze, Davighi:2021osh},
non-linear bounds from moment theorems \cite{Bellazzini:2020cot, Arkani-Hamed:2020blm, Chiang:2021ziz} and exploiting full crossing symmetry \cite{Tolley:2020gtv, Caron-Huot:2020cmc, Sinha:2020win, Raman:2021pkf, Haldar:2021rri},
generalised bounds for EFTs with multiple fields \cite{Li:2021cjv, Du:2021byy}, and bounds including the effects of gravity \cite{Alberte:2020bdz, Alberte:2020jsk, Tokuda:2020mlf, Herrero-Valea:2020wxz, Caron-Huot:2021rmr, Alberte:2021dnj}.
}. 
By using these bounds, we can carve the EFT parameter space into regions which could admit a standard Wilsonian UV completion, and regions which could \emph{never} be embedded in a high-energy theory (in a way consistent with these standard axioms).
This positivity technology has recently been applied to a number of different low-energy EFTs, ranging from particle physics \cite{Pham:1985cr,Ananthanarayan:1994hf,Pennington:1994kc,Distler:2006if,Vecchi:2007na, Bellazzini:2017bkb, Bellazzini:2018paj,Zhang:2018shp,Bi:2019phv, Remmen:2019cyz, Englert:2019zmt, Yamashita:2020gtt, Remmen:2020vts, Bonnefoy:2020yee, Trott:2020ebl, Chala:2021wpj} 
to cosmology 
\cite{Bellazzini:2015cra,Cheung:2016yqr, Bonifacio:2016wcb, Bellazzini:2017fep, deRham:2017xox, deRham:2018qqo, Alberte:2019xfh, Alberte:2019zhd, Wang:2020xlt, Bonifacio:2018vzv, Bellazzini:2019bzh, deRham:2017imi,Herrero-Valea:2019hde, Bellazzini:2019xts}. 
However, these previous applications have focussed on scattering fluctuations around a trivial vacuum configuration, such as $\phi = 0$.  

~

In this work, we develop a new application of EFT positivity bounds.
Rather than focus on the scaterring processes around the trivial (background) solution, $\phi = 0$, which most related work has focused on in the past, we also consider $2\to2$ scattering processes around other stable vacua of the EFT---this generates a family of different positivity bounds, which can be used to diagnose which vacua may persist in the UV completion.  
This is especially relevant in the dark energy context, where large classes of observationally relevant scalar-tensor theories  do not permit a stable $\phi = 0$ solution---see \cite{Barreira:2013jma,Deffayet:2010qz,Traykova:2021hbr} and references therein.
In this paper, to illustrate the key concepts as simply as possible, we focus on the covariant Galileon theory \cite{Deffayet:2009wt}, 
\begin{eqnarray}\label{CovGal_action}
S [\phi ]=\int d^4x \sqrt{-g}\left\{\frac{\MPl^2}{2}R+\sum_{i=1}^5 c_i {\cal L}_i[ \phi ]\right\} \, ,
\end{eqnarray}
which is characterised by five constant (dimensionless) Wilson coefficients, $c_i$. 
The five terms in the Lagrangian are given by
\begin{align}
\mathcal{L}_1&= -\frac{1}{2} \Lambda_3^3 \phi ~,\quad\quad  \mathcal{L}_2 =  X ~,\quad\quad \mathcal{L}_3 = 2 X[\Phi] ~, \nn \\ \mathcal{L}_4 &= \frac{X^2}{\Lambda_3^6} R + 2 X \left( [\Phi]^2-[\Phi^2] \right) ~,\nn \\
\mathcal{L}_5&= {\frac{X^2}{\Lambda_3^6}G_{\mu\nu}\Phi^{\mu\nu}} - \frac{X}{3} \left( [\Phi]^3 + 2[\Phi^3] -3[\Phi][\Phi^2] \right).
\label{lags}
\end{align}
where $[\Phi^n]$ are traces of the matrix $ \Phi_{\mu}^{\; \nu} \equiv   \nabla_\mu \nabla^\nu\phi / \Lambda_3^3$, $X \equiv -\tfrac{1}{2} g^{\mu\nu}\nabla_\mu\phi\nabla_\nu\phi$, $G_{\mu\nu}$ is the Einstein tensor, and $\Lambda_3$ is the strong coupling scale of this EFT. 
Note that the Galileon invariance of the pure scalar interactions in this theory is broken by gravitational corrections, but since graviton exchange is suppressed by at least one factor of $\MPl$ the above remains radiatively stable---see \cite{Luty:2003vm,Nicolis:2004qq,deRham:2010eu,Burrage:2010cu,deRham:2014wfa, Pirtskhalava:2015nla} 
and also \cite{deRham:2012ew,Goon:2016ihr,Saltas:2016nkg,Noller:2018eht,Heisenberg:2019wjv,Heisenberg:2020cyi,Goon:2020myi} for recent extensions of these arguments (also in other cosmologically relevant scalar-tensor theories).
We emphasise that this example is not chosen to champion this specific theory as a frontrunner for e.g. dark energy, but rather since (as we will see) it neatly illustrates a number of key conceptual points related to positivity bounds in theories where multiple vacua exist and also allows us to explore the interface of such bounds with observational constraints in detail. 

~

In particular, for any specific choice of the $c_n$, this theory has up to four stable Lorentz-invariant solutions of the form $\phi \propto\,   x_\mu x^\mu$, plus various additional solutions which break Lorentz boosts. 
By applying positivity bounds around each of the stable vacua of this theory, we find that usually only a subset are consistent with unitarity, causality \blue{(analyticity\footnote{
When considering vacua which spontaneously break Lorentz symmetry, the low-energy positivity bounds that we consider correspond to an analyticity in UV which can be stronger than that implied by causality \cite{Baumann:2015nta, Grall:2021xxm}.
})} and locality in the UV.  
This illustrates the importance of the IR vacuum structure when it comes to searching for possible UV completions, and provides a more sophisticated ``positivity-map'' of parameter space in which different regions correspond to the UV containing different stable vacua. 
For instance, we show that while the quartic Galileon cannot be generated with $c_4 > 0$ if the UV theory contains the trivial $\phi =0$ vacuum, it \emph{can} be generated with a positive sign if the kinetic term has $c_2 < 0$ and only the non-trivial vacuum is stable.
That a non-trivial vacuum structure may allow for an apparent violation of the $\phi = 0$ positivity bounds was also recently observed in \cite{Davis:2021oce, Aoki:2021ffc} for $P(X)$ theories. 

~

These considerations are particularly important once we pair the theoretical positivity considerations above with constraints from cosmological observations. Indeed, for Galileon dark energy cosmological observations require that $c_2 < 0$ \cite{Barreira:2013jma} and so the trivial $\phi = 0$ solution is always unstable in realistic such models. This conclusion remains true in a much wider class of scalar-tensor theories \cite{Traykova:2021hbr}, so the Galileon cosmologies we consider here are representative in this sense.
It is therefore not possible to apply conventional positivity bounds about the trivial vacuum for these theories, and instead we must only consider bounds from the stable non-trivial vacua. These bounds can then be incorporated into a cosmological constraints analyses as theoretical priors along the lines explored in \cite{Melville:2019wyy, deRham:2021fpu}. 
Observational constraints for cosmological Galileons have been well explored  \cite{Barreira:2013jma,Okada:2012mn,Bellini:2013hea,Barreira:2014zza,Barreira:2014jha,Burrage:2015lla,Neveu:2016gxp,Renk:2017rzu,Leloup:2019fas}, so in addition to its interesting vacuum structure this further motivates choosing the Galileon as an initial example to investigate the interplay between novel positivity bounds and observational constraints. 
Here we therefore explore for the first time how demanding the existence of different vacua places different positivity priors on cosmological parameter spaces. 
For instance, 
we identify an incompatibility between certain vacua in the EFT and large values of the so-called ``tracking parameter'' $\xi$, that parametrises the background evolution of the dark energy Galileon. 
We also identify a particular boost-breaking vacuum which is (at least instantaneously) very close to the cosmological evolution, such that a violation of the corresponding positivity bounds would be difficult to reconcile with a standard UV completion. Positivity bounds computed around this vacuum are therefore particularly well-motivated theoretical priors in a cosmological setting and we show that this prior is indeed powerful, e.g. ruling out the vast majority of otherwise observationally viable cosmologies with $c_3 > 0$. 

~

In Section~\ref{sec:vacua}, we describe various vacua (background solutions) of the covariant Galileon \eqref{CovGal_action} and the conditions under which they are stable. 
Then in Section~\ref{sec:positivity} we consider $2\to2$ scattering of fluctuations about each of these vacua and derive the corresponding positivity bounds.
In Section~\ref{sec:observations} we discuss the constraints on~\eqref{CovGal_action} from cosmological observations, and finally in Section~\ref{sec:priors} we compare each of the different positivity bounds with observations and discuss their use as theoretical priors. 
We conclude with some discussion in Section~\ref{sec:discussion}. Further technical details, especially on the (derivation of the) positivity bounds employed in the main text, can be found in the Appendices.

\section{A study of IR vacua: From Galileon to Galileid}
\label{sec:2}

\subsection{Vacua, fluctuations and stability}
\label{sec:vacua}

In this subsection, we describe various background solutions for the scalar field $\phi$ in the covariant Galileon theory~\eqref{CovGal_action}. 
We begin by considering a non-dynamical flat spacetime background, $g_{\mu\nu} = \eta_{\mu\nu}$ (and hence set $R = 0 = G_{\mu\nu}$ in \eqref{lags}, which reduce to the usual Galileon interactions). 
The coupling to gravity will be re-introduced when we discuss cosmological backgrounds in Section \ref{sec:priors}, but note that since they play no role in this section the corresponding results apply to any covariantisation of the Galileon interactions.

\paragraph{Background:}
We investigate a family of background solutions which are specified by two constant parameters: $\alpha$ and $\beta$. Specifically, we expand the scalar $\phi = \bar{\phi} + \varphi$ about a background configuration $\bar{\phi}$ given by,
\begin{align}
\bar{\phi} = \frac{\Lambda_3^3}{2} \left( - \alpha t^2 + \beta | \mathbf{x} |^2 \right) \; .
\label{eqn:galileid}
\end{align}
Since the action \eqref{CovGal_action} is invariant under a shift and (on a fixed flat spacetime) a Galileon symmetry, $\phi \to \phi + c + c_\mu x^\mu$, the background~\eqref{eqn:galileid} depends on the lowest power of the co-ordinates which cannot be removed using this symmetry.  
Three subcases are of particular interest for us:
\begin{itemize}
\item $\alpha = \beta = 0$. This case corresponds to the usual trivial vacuum $\bar \phi  = 0$.
\item $\alpha = \beta$. This case encapsulates the trivial vacuum above, but also includes non-trivial, but still Lorentz invariant, vacua of the form $\bar\phi \propto x_\mu x^\mu$. We therefore have a one-parameter family of Lorentz invariant background solutions.
\item $\alpha \neq \beta$. This two-parameter background solution breaks boosts, but a combination of the Galileon symmetry and spacetime translations survives\footnote{
Note that while the stress-energy tensor $T_{\mu\nu} \propto x^2$ is not translationally invariant, we are working in the limit where gravity has decoupled and so the spacetime background remains approximately Minkowski.
} and so the effective interactions for $\varphi$ conserve both energy and momentum. This produces the so-called \emph{type-II galileid} studied in \cite{Nicolis:2015sra}\footnote{
The \emph{type-I galileid} corresponds to $\alpha = 0$.
}.
\end{itemize}
The solution \eqref{eqn:galileid} encapsulates all of the above subcases and we will refer both to this general solution as well as to the three subcases above throughout this paper.
Expanding $\phi$ about $\bar{\phi}$ in $S [ \phi ]$ leads to an effective action for the $\varphi$ fluctuations, and in the remainder of this section we describe this action for each of the above backgrounds. 

\paragraph{Fluctuations about trivial background:}
For the trivial background $\bar \phi = 0$, $\varphi$ is governed by the same covariant Galileon action $S[\varphi]$ given in~\eqref{CovGal_action}. 
In order for this to be a stable background, the tadpole term must vanish, $c_1=0$, and the kinetic term must have the sign $c_2 > 0$. 
To canonically normalise the fluctuations, we should rescale $\varphi \to \varphi / \sqrt{c_2}$ (i.e. the effective strength of the cubic Galileon interaction is $c_3 c_2^{-3/2}$, as measured by any observable such as a scattering amplitude).

\paragraph{Fluctuations about other Lorentz-invariant backgrounds:}
If we instead consider the general Lorentz-invariant background, $\bar{\phi}_{\rm LI} = \beta \Lambda_3^3 x^\mu x_\mu/2$, then the $\varphi$ fluctuations will be described by an effective action of the same Galileon form, but with redressed coefficients,
\begin{align}
\bar{c}_1 &=  c_1 - 8 \beta c_2 -48 \beta^2 c_3 - 96 \beta^3 c_4 + 16 \beta^4 c_5  \nonumber \\
\bar{c}_2 &= c_2 + 12 \beta c_3 + 36 \beta^2 c_4 - 8 \beta^3 c_5 \nonumber \\
\bar{c}_3 &= c_3 + 6 \beta c_4 - 2 \beta^2 c_5 \label{eqn:coefs1}    \\
\bar{c}_4 &= c_4 - \frac{2}{3} \beta c_5  \nonumber \\
\bar{c}_5 &=  c_5 \; , \nonumber 
\end{align}
since schematically each $c_n \partial^{2n-2} \phi^n$ interaction in \eqref{CovGal_action} produces $\sum_{j=0}^n \beta^{n-j} \partial^{2j-2} \varphi^{j}$ once expanded about this background.
This background is stable whenever $\bar{c}_1 = 0$ and $\bar{c}_2 > 0$. Note that in any particular covariant Galileon theory (i.e. a particular choice of the $c_n$ in \eqref{CovGal_action}), there are up to four real background solutions for $\beta$. 
Canonically normalising the fluctuations corresponds to rescaling $\varphi \to \varphi / \sqrt{\bar{c}_2}$ (so overall the effective strength of the cubic Galileon interaction is now $\bar{c}_3 \bar{c}_2^{-3/2}$, and has a non-polynomial dependence on $\beta$).

\paragraph{Fluctuations about Lorentz-breaking backgrounds:}
Finally we turn to the general case of \eqref{eqn:galileid} with $\alpha \neq \beta$. 
This can be written as $\bar{\phi} = \bar{\phi}_{\rm LI} + (\beta - \alpha )  \Lambda_3^3 t^2/2$, and we see that the difference $\beta-\alpha$ controls the breaking of boost invariance. 
The classical stability of \eqref{eqn:galileid} is determined by the linear and quadratic terms,
\begin{align}
\delta S [\varphi] &=  \int dt d^3 \mathbf{x} \; \left(  -  \frac{1}{2} J  \varphi \right)   \nonumber \\
\delta^2 S [\varphi] &= \int dt d^3 \mathbf{x} \; \frac{Z^2}{2 c_s^3} \left( \dot \varphi^2 - c_s^2 ( \partial_i \varphi )^2 \right)  \; , 
\label{d2S}
\end{align}
where the tadpole and kinetic-term normalisation are shifted relative to \eqref{eqn:coefs1} by the boost-breaking part of $\bar\phi$,
\begin{align}
\label{Jcs}
 J = \bar{c}_1 + 2 \bar{c}_2 ( \beta - \alpha  )  \; , \;\;\; 
  Z^2/c_s^3 = \bar{c}_2  \; , \;\;\;  
 1 - c_s^2  = (\beta - \alpha) \frac{4 \bar{c}_3}{\bar{c}_2} \; .
\end{align}
and now the sound speed $c_s$ of these fluctuations is generally different from $1$ (the invariant speed preserved by Lorentz boosts).
In order for this to be a stable background, we require that the tadpole vanishes, $J=0$ and that there are no ghosts or gradient instabilities, $Z^2> 0$ and $c_s^2 > 0$. 
To canonically normalise the fluctuations, we should rescale $\varphi \to \tilde{\varphi} =  Z \varphi$, so that the propagation of $\tilde{\varphi}$ fluctuations is determined by the simple effective metric $\tilde{g}^{\mu\nu} = \text{diag} \left( -1 , c_s^2 , c_s^2 , c_s^2  \right)$, i.e. $\delta^2 S [ \varphi ] = \int dt d^3 \mathbf{x} \sqrt{-\tilde{g}} \, \left( - \frac{1}{2} \tilde{g}^{\mu\nu} \partial_\mu \tilde{\varphi} \partial_\nu \tilde{\varphi} \right)$.

In addition to the non-trivial sound speed, the breaking of boosts also allows for new interactions.
The cubic and quartic action for the fluctuations can be written as,
\begin{align}
\delta^3 \mathcal{S} [ \varphi ] &= 
\int dt d^3 \mathbf{x} \sqrt{-\tilde{g}}  \; 
 \left( 
- \tilde{c}_{3}   ( \tilde{\partial} \varphi )^2
- \tilde{d}_{3}  \dot{ \varphi }^2 \right) \frac{ \tilde{\Box} \varphi }{  \Lambda_3^3 }
  \label{eqn:Scubic}  \\
\delta^4 \mathcal{S} [ \varphi ] &= 
\int dt d^3 \mathbf{x} \sqrt{-\tilde{g}}  \; 
 \left( 
- \tilde{c}_{4}  ( \tilde{\partial}  \varphi )^2
- \tilde{d}_{4}  \dot{ \varphi }^2  \right) \frac{ ( \tilde{\Box}  \varphi )^2 -  (\tilde{\partial} \tilde{\partial}  \varphi )^2 }{   \Lambda_3^6 }
    \label{eqn:Squartic}
\end{align}
where $\sqrt{{-\tilde{g}}} = 1/c_s^3$ and these derivatives are contracted using the effective metric\footnote{
In particular, note that if $n^\mu = \nabla^\mu t$ is the time-like direction, then $c_s^2 \eta^{\mu\nu} = \tilde{g}^{\mu\nu} + (1- c_s^2) n^\mu n^\nu$, which allows us to replace any $\partial \varphi$ with $ \tilde{\partial} \varphi$ and $\dot \varphi$.
}, $\tilde{\partial}^\mu = \tilde{g}^{\mu\nu}  \partial_\nu$ (e.g. $\tilde{\Box} = \tilde{g}^{\mu\nu} \partial_\mu \partial_\nu$). 
The interaction coefficients are given by\footnote{
For convenience, in Appendix~\ref{app:galileid} we provide expressions for the effective coefficients ($J$, $Z^2$, $c_s^2$, $\tilde{c}_n$, $\tilde{d}_n$) in terms of the original $c_n$ coefficients appearing in \eqref{CovGal_action}.
}, 
\begin{align}
\sqrt{- \tilde{g}} \, \tilde{c}_3 &= \frac{1}{c_s^4} \left[  \bar{c}_3 - 3 \bar{c}_4 ( \beta - \alpha )  \right] \; , 
&\sqrt{- \tilde{g}} \, \tilde{d}_3 &= 2 \frac{\beta - \alpha }{c_s^4} \left[  \frac{ 4 \bar{c}_3^2 }{\bar{c}_2} - 3 \bar{c}_4   \right]  \nonumber \\
\sqrt{-\tilde{g}} \, \tilde{c}_4 &= \frac{1}{c_s^6} \left[  \bar{c}_4 + \frac{2}{3} \bar{c}_5 ( \beta - \alpha )  \right]  \; ,
&\sqrt{-\tilde{g}} \, \tilde{d}_4 &= 3 \frac{\beta - \alpha}{c_s^6} \left[  \frac{4 \bar{c}_3 \bar{c}_4}{\bar{c}_2} + \frac{2}{3} \bar{c}_5   \right]   
\label{eqn:coefs2}
\end{align}
and clearly reduce to \eqref{eqn:coefs1} when $\alpha \to \beta$, the limit in which $c_s \to 1$ and Lorentz symmetry is restored. 
Further, the coefficients \eqref{eqn:coefs2} differ from \eqref{eqn:coefs1} by at most one factor of $(\beta - \alpha)$, which reflects the antisymmetric derivative structure of the Galileon interactions (i.e. at most one leg can be put on the $(\beta- \alpha)t^2$ part of the $\bar\phi$ background). 
Crucially, it is the effective couplings \eqref{eqn:coefs2} that control the strength of $\tilde{\varphi}$ interactions on this $\bar\phi$ background, and therefore it is these particular combinations of parameters which will appear in the positivity bounds.

\subsection{Positivity bounds}
\label{sec:positivity}

Having described the vacua in this low-energy EFT, we now turn to the question of UV completion. 
Positivity bounds provide a way of diagnosing whether a particular EFT can ever be embedded into a consistent UV complete theory (i.e. one which is unitary, causal and local) in the usual Wilsonian sense.
In this section, we derive a variety of positivity bounds for the covariant Galileon EFT \eqref{CovGal_action}, and show that they can be used to determine which of the EFT vacua (i.e. which values of $\alpha$ and $\beta$) can remain viable backgrounds in the full UV theory. 

\paragraph{Positivity about trivial background:}
First, let us consider the trivial background $\bar \phi = 0$ (i.e. $\beta = \alpha = 0$).
The $2 \to 2$ scattering amplitude for the Galileon interactions about this background, and the corresponding positivity bounds, can be found in \cite{Adams:2006sv, deRham:2017imi}.  
In this case, Lorentz invariance is unbroken, and consequently the amplitude is a simple function of two Mandelstam variables only, $\mathcal{A} (s,t)$. 
At tree-level, this EFT amplitude can be written as an analytic power series in $s$ and $t$, which in particular contains the terms,
\begin{align}
\mathcal{A} (s,t) = c_{ss} s^2 + c_{sst} s^2 t + ... \; ,
\label{eqn:Aexp}
\end{align}
for constant coefficients $c_{ss}$ and $c_{sst}$. 
This EFT admits a unitary, causal, local, Lorentz invariant UV completion only if \cite{Adams:2006sv},
\begin{align}
 c_{ss} > 0 \; , 
 \label{eqn:posLO_css}
\end{align}
and \cite{deRham:2017avq}
\begin{align}
 c_{sst} > - \frac{3}{2 \Lambda^2} c_{ss} \; , 
 \label{eqn:posNLO_csst}
\end{align}
where $\Lambda$ is the cut-off of the EFT. The technical steps leading from unitarity, analyticity and locality in the UV to these constraints in the EFT are reviewed briefly in Appendix~\ref{app:positivity}. 

An exact Galileon symmetry requires that $c_{ss} = 0$, which violates the first of these bounds \cite{Adams:2006sv}. 
One possible resolution is to break the Galileon symmetry, e.g. with a small mass \cite{deRham:2017imi} or small $(\partial \phi)^4$ interaction \cite{Bellazzini:2017fep}, but this inevitably leads to an unacceptably low cut-off\footnote{
In particular, exploiting full crossing symmetry gives a further bound \cite{Tolley:2020gtv}, $c_{sst} < +8 c_{ss} / \Lambda^2$, and so it is not possible to arrange for a small $|c_{ss}| \ll \Lambda^2 |c_{sst}|$, i.e. the symmetry breaking must be order-one in units of the cut-off. 
}. 
Another possible resolution is to make weaker assumptions about the nature of the UV completion---for instance, if the UV amplitude exhibits some mild non-locality,
then the dispersion relation for $\partial_s^2 \mathcal{A}$ need not converge and \eqref{eqn:posLO_css} need not apply, and yet higher order positivity bounds may still be used to constrain these couplings \cite{Keltner:2015xda, Davighi:2021osh}. 
Yet another possibility is that gravitational effects modify \eqref{eqn:posLO_css}, particularly in light of recent results which show that a small negative $c_{ss}$ may not violate causality if balanced by a gravitational time delay \cite{Alberte:2020bdz} (see also \cite{Hollowood:2015elj, deRham:2019ctd, deRham:2020zyh, deRham:2021bll, Chen:2021bvg, Arkani-Hamed:2021ajd}).
Since our goal here is to compare positivity bounds around different vacua, rather than rescue any particular covariant Galileon theory, we will accept that the leading bound \eqref{eqn:posLO_css} appears to be marginally violated for the trivial background considered here and move on to contemplate the first non-trivial bound \eqref{eqn:posNLO_csst}. 

The bound on $c_{sst}$ has recently been used to constrain Horndeski scalar-tensor theories \cite{Melville:2019wyy, deRham:2021fpu}, where it was shown that \eqref{eqn:posNLO_csst} provides a potentially useful theoretical prior when paired with current observational data. 
For the covariant Galileon \eqref{CovGal_action} (i.e. a particular choice of the Horndeski functions), this bound is simply \cite{deRham:2017imi},
\begin{align}
c_3^2 - c_2 c_4 > 0 \; . 
\label{eqn:pos_trivial}
\end{align}
Physically, this positivity condition is telling us that it is not possible to integrate out unitarity, causal, local UV physics to produce an EFT of the form \eqref{CovGal_action} with a stable $\bar \phi = 0$ background ($c_1 = 0$ and $c_2 > 0$) and with $c_3^2 < c_2 c_4$. 
This has a number of important consequences---for instance, in the absence of the cubic Galileon, $c_3 = 0$, the region $c_4 > 0$ is forbidden, and this is the precisely the region in which Vainshtein screening takes place~\cite{Dvali:2012zc, Davis:2021oce}. 

In this work, we are exploring for the first time the caveat that the above argument implicitly assumes $\bar \phi = 0$ is a stable background solution about which to compute $\mathcal{A} (s,t)$. 
For example, if $c_2 < 0$ then the action \eqref{CovGal_action} expanded around $\bar\phi = 0$ should not be used to compute a perturbative scattering process: rather, one must first expand around a solution $\bar{\phi}$ which is stable, and consider scattering $\varphi$ fluctuations using the effective action given above in~\eqref{eqn:Scubic} and \eqref{eqn:Squartic}. 
Such cases are not just a calculational curiosity, but highly relevant in cosmological contexts. As we shall discuss in more detail below, $c_2 < 0$ is observationally mandated in the context of the self-accelerating Galileon dark energy theories we are focusing on here~\cite{Barreira:2013jma} and indeed this conclusion remains true in a much wider classes of observationally relevant scalar-tensor theories~\cite{Traykova:2021hbr}. A stable $\bar \phi = 0$ does not exist in these theories and hence arguments and results based on assuming the existence of such a background solution, e.g. the positivity requirement~\eqref{eqn:pos_trivial}, do not apply in this context.

\paragraph{Positivity about other Lorentz-invariant backgrounds:}
To make this more concrete, consider what happens when one of the $\alpha = \beta \neq 0$ vacua is stable. 
This background preserves Lorentz invariance and (thanks to the Galileon symmetry) translation invariance, so the amplitude $\mathcal{A}_{\varphi \varphi \to \varphi \varphi} (s,t)$ is once again a function of $s$ and $t$ only.
We give this function explicitly in Appendix~\ref{app:positivity}.
The positivity bound \eqref{eqn:posNLO_csst} applied to this amplitude gives,
\begin{align}
 \bar{c}_3^2 - \bar{c}_2  \bar{c}_4 > 0 \; , 
 \label{eqn:pos_LI}
\end{align}
where the $\bar{c}_n$ coefficients are given in \eqref{eqn:coefs1}. 
Although we have adopted a notation in which \eqref{eqn:pos_trivial} and \eqref{eqn:pos_LI} are superficially similar, let us stress that the bound \eqref{eqn:pos_LI} about this $\bar \phi_{\rm LI} = \tfrac{1}{2} \beta \Lambda_3^3 \,  x_\mu x^\mu$ background is qualitatively different when $\beta \neq 0$. 
For one thing, while the quintic coefficient $c_5$ of the original Lagrangian \eqref{CovGal_action} could never be constrained by tree-level scattering about $\bar \phi = 0$, considering this non-trivial background has introduced a $c_5$ dependence into \eqref{eqn:pos_LI}\footnote{
The use of non-trivial backgrounds to constrain higher-point interactions has been described previously in \cite{Davis:2021oce}, where it was shown that for simple $P(X)$ theories this strategy can reproduce the bounds that would have been inferred from higher-point scattering amplitudes~\cite{Chandrasekaran:2018qmx}.
This is analogous to the observation that linearising interactions around several different backgrounds can probe information that only enters non-linearly around a specific background.
}. 
A further interesting observation is that when $c_3 = c_5 = 0$ and one considers the quartic Galileon interaction alone, \eqref{eqn:pos_LI} becomes $-c_2 c_4 > 0$ for \emph{any} value\footnote{
The fact that $\beta$ drops out of the positivity bound is perhaps related to the special Galileon symmetry $\phi \to \phi + S^{\mu\nu} \left( x_\mu x_\nu  - \frac{c_4}{c_2} \partial_\mu \phi \partial_\nu \phi \right)$ enjoyed by the quartic Galileon, where $S^{\mu\nu}$ is a fixed traceless symmetric tensor \cite{Cheung:2014dqa, Hinterbichler:2015pqa}.
} of $\beta$. However, now the condition for stability is $c_2 + 36 \beta^2 c_4 > 0$, and so unlike the bound \eqref{eqn:pos_trivial} around the trivial background we now have the possibility of setting $c_2 < 0$ and $c_4 > 0$. Put another way, the quartic interaction in \eqref{CovGal_action} could be generated with a positive sign by unitary, causal, local UV physics if $c_2 < 0$ and the stable background is $\bar \phi =\tfrac{1}{2}  \beta \Lambda_3^3 \, x_\mu x^\mu$ (with $\beta^2 = - c_2/(12 c_4)$ from $J=0$), but not if $c_2 > 0$ and $\bar \phi=0$ is the stable background. 
This will have interesting consequences once we consider the interplay with observational bounds on the $c_i$ in the next section (particularly in the observation-favoured $c_2 < 0$ regime alluded to above).

\paragraph{Positivity about Lorentz-breaking backgrounds:}
Finally, we consider backgrounds of the form~\eqref{eqn:galileid} in which $\alpha \neq \beta$. 
\blue{Importantly, here $c_2 < 0$ is perfectly consistent with the existence of stable background solutions -- this follows directly from \eqref{Jcs} -- so the observational constraints mandating a negative $c_2$ for cosmological solutions mentioned above do not rule out the existence of stable $\{\alpha,\beta\}$ vacua within the same EFT. While this by itself does not guarantee that cosmological and galileid solutions co-exist within the same EFT, this removes a major obstacle in linking (positivity) bounds and (observational) constraints from those two classes of vacua. Indeed, later on we will discuss how precisely one may map constraints from the galileid vacua considered here to the cosmological ones detailed in section \ref{sec:3}.} 

For $\alpha \neq \beta$ backgrounds, the amplitude for scattering $\varphi$ fluctuations is no longer invariant under boosts. $\mathcal{A}$ can therefore depend explicitly on three additional variables, which we take to be the energies $\{ \omega_1, \omega_2 ,  \omega_3 \}$ of the fluctuations (since time translations are unbroken $\omega_4 = - \omega_1 - \omega_2 - \omega_3$ is fixed by energy conservation).
In the forward limit, in which $t=0$ and $\omega_3 = - \omega_1$, the most general boost-breaking amplitude takes the form, 
\begin{align}
\mathcal{A}  = c_{ss} s^2 + d_{s\omega \omega} s \omega_1 \omega_2 + d_{\omega\omega\omega\omega} \omega_1^2 \omega_2^2 + ... 
\label{eqn:ABoostlessFwd}
\end{align}
plus terms that are suppressed either by more powers of the cut-off or by the approximate shift symmetry of the fields (as we show in Appendix~\ref{app:noGalSymm}).

The analogue of the positivity bound \eqref{eqn:posLO_css} for such an EFT amplitude \eqref{eqn:ABoostlessFwd} was first worked out in~\cite{Baumann:2015nta}, and amounts to fixing the energies as though in the centre-of-mass frame (i.e. $\omega_1 = \omega_2 = \sqrt{s}/2$), which gives $\mathcal{A}$ an effective $s^2$ coefficient that should be positive,
\begin{align}
 c_{ss} + \frac{1}{4} d_{s \omega \omega} + \frac{1}{16} d_{\omega\omega\omega\omega} > 0 \; ,
\label{eqn:posLO_css2}
\end{align}
if the UV completion obeys properties analogous to unitarity, causality and locality---namely it is smoothly connected to the EFT amplitude modulo unitary branch cuts (we briefly review these technical details in Appendix~\ref{app:derivation}). 
More recently, it was pointed out in \cite{Grall:2021xxm} that for general particle energies (i.e. scattering with a general centre-of-mass momentum $\mathbf{p}_1 + \mathbf{p}_2 \neq 0$) the same UV axioms lead to a 1-parameter family of bounds,
\begin{align}
 c_{ss} + \frac{\gamma^2}{4} d_{s \omega \omega} + \frac{\gamma^4}{16} d_{\omega\omega\omega\omega} > 0 \; ,
 \label{eqn:posLO_css3}
\end{align}
controlled by the dimensionless ratio $\gamma^2 = 4 \omega_1 \omega_2 /s$ which may take any value $\geq 1$. Setting $\gamma^2=1$ recovers \eqref{eqn:posLO_css2}. Physically, $\gamma$ is related to (the Lorentz factor of) the centre-of-mass velocity: we describe this interpretation in more detail in Appendix~\ref{app:gamma}. 

For the general galileid background~\eqref{eqn:galileid}, we find that $c_{ss} = d_{s \omega\omega} = d_{\omega \omega \omega \omega} = 0$ in \eqref{eqn:ABoostlessFwd}, i.e. the Galileon symmetry forbids these terms from appearing in the amplitude. Consequently, \eqref{eqn:posLO_css3} is always marginally violated, for any value of $\alpha$ and $\beta$. 
This answers a question first posed in \cite{Nicolis:2015sra}: the Galileid suffers from the same IR obstruction to UV completion as the Galileon, which cannot be healed by considering non-trivial backgrounds (at least those which preserve time-translation invariance).
However, note that this marginal violation of the leading positivity bound \eqref{eqn:posLO_css3} is again subject to the same possible loopholes as the Lorentz-invariant bound \eqref{eqn:posLO_css} described above.
Ultimately, since our goal is to compare positivity bounds around different vacua, rather than rescue any particular covariant Galileon theory from \eqref{eqn:posLO_css3}, we will move on to the first non-trivial bound which arises beyond the forward limit. 

Allowing for a small non-zero $t$, the most general boost-breaking amplitude compatible with a Galileon symmetry takes the form\footnote{
Note that we continue to set $\omega_1 = -\omega_3$ so that $t=0$ corresponds to the forward limit $p_1 = -p_3$ (and $p_2 = -p_4$)---this ensures the positivity of the UV contribution to the dispersion relation, from which \eqref{eqn:posNLO_csst2} follows (see Appendix~\ref{app:derivation}).
For the sake of completeness, the amplitude at general $\omega_1 \neq \omega_3$ is given in~\eqref{eqn:A}.    
},
\begin{align}
\mathcal{A} \left( s, \;  t,  \; \omega_1 = -\omega_3 , \; \omega_2  \right) &= c_{sst} s^2 t 
+ d_{s t \omega \omega} s t \omega_1 \omega_2 + ... 
\label{eqn:Aexp2}
\end{align}
plus higher order terms, as we show in Appendix~\ref{app:noGalSymm}. 
In terms of the Wilson coefficients appearing in \eqref{eqn:Scubic} and \eqref{eqn:Squartic}, 
\begin{align}
c_{sst} &= 3 \left( \frac{\tilde{c}_3^2}{Z^2}   - \tilde{c}_4 \right)   \; , &d_{st \omega\omega} &= 4 \left( \tilde{d}_4 - \frac{2 \tilde{c}_3 \tilde{d}_3}{Z^2} \right) \; . 
\end{align}
Positivity bounds for such an EFT amplitude were \blue{recently derived in \cite{Grall:2021xxm} by assuming an analytic extension of $\mathcal{A}$ into the UV consistent with unitarity and locality, which in this case would require}\footnote{
Note that, in the absence of any Galileon symmetry, there can be additional terms appearing in the amplitude \eqref{eqn:Aexp2} which affect \eqref{eqn:posNLO_csst2}---we give the general bound on any shift-symmetric amplitude in Appendix~\ref{app:noGalSymm}.
},
\begin{align}
 c_{sst} + \frac{\gamma^2}{4} d_{st \omega\omega} > 0 \; ,
 \label{eqn:posNLO_csst2}
\end{align}
(see Appendix~\ref{app:positivity} for a brief review of the technical details).
\blue{In this work, we focus for the most part on the consequences of this bound \eqref{eqn:posNLO_csst2}, which can be applied to a greater range of vacua than the Lorentz-invariant bound \eqref{eqn:posNLO_csst}.}
The analogue of the positivity bound \eqref{eqn:posNLO_csst} corresponds to setting $\gamma = 1$ (i.e. scattering with ``centre-of-mass'' kinematics), and gives the following constraint on the Wilson coefficients, 
\begin{align}
\text{Pos. bound 1:}  \qquad\qquad
\tilde{c}_3^2  - Z^2  \tilde{c}_4 > \frac{1}{3} \left( 2 \tilde{c}_3 \tilde{d}_3 -  Z^2  \tilde{d}_4 \right) \; .
\label{eqn:dt_pos1}
\end{align}
Note that when $\alpha = \beta$ and the $\phi$ background recovers Lorentz-invariance, the coefficients $\tilde{d}_3$ and $\tilde{d}_4$ on the right-hand-side vanish and we recover the Lorentz-invariant bound~\eqref{eqn:pos_LI}. 
Furthermore, one key conceptual difference to the Lorentz-invariant case is that now the positivity bound \eqref{eqn:posNLO_csst2} depends on the additional parameter, $\gamma^2$. 
By considering large values of $\gamma$ (i.e. scattering processes in which there is a non-zero centre-of-mass motion\footnote{
We show in Appendix~\ref{app:unitarity} that $\gamma \gg 1$ at small $s$ remains within the validity of the EFT, and further discuss this interesting limit in Appendix~\ref{app:gamma}. 
}), there is an \emph{additional} constraint,
\begin{align}
\text{Pos. bound 2:}  \qquad\qquad
2 \tilde{c}_3 \tilde{d}_3 - Z^2  \tilde{d}_4  \leq 0  \; .
\label{eqn:dt_pos2}
\end{align}
Taken together, \eqref{eqn:dt_pos2} and \eqref{eqn:dt_pos1} imply a \emph{weaker} bound on $\tilde{c}_3^2 - Z^2 \tilde{c}_4$ than the Lorentz invariant~\eqref{eqn:pos_LI}. 
In terms of these effective coefficients, allowing for a more general vacuum/background for $\phi$ therefore allows a greater range of Wilson coefficients to be UV completed\footnote{
While this is certainly intuitive, note that when written in terms of the original $c_n$ coefficients appearing in the covariant Lagrangian~\eqref{CovGal_action}, the effective $\tilde{c}_n$ are themselves functions of $\alpha$ and $\beta$ (see \eqref{eqn:coefs2}), and so \eqref{eqn:dt_pos1} for a boost-breaking background is not necessarily a weaker positivity bound on the $c_n$ parameter space. 
}. When comparing with observational constraints in section~\ref{sec:3}, we will find that \eqref{eqn:dt_pos1} and \eqref{eqn:dt_pos2} taken together are significantly more constraining than \eqref{eqn:dt_pos1} alone.

\blue{It should be noted that the Lorentz-invariant bound \eqref{eqn:pos_LI} and the boost-breaking bounds (\ref{eqn:dt_pos1}, \ref{eqn:dt_pos2}) are on a slightly different footing. While both require the UV completion to be analytic in the complex $s$-plane, in the case of \eqref{eqn:pos_LI} this analyticity is a direct consequence of causality (i.e. local operators commute outside of the light-cone). For backgrounds which spontaneously break boosts, the connection between causality and analyticity is more subtle (in particular the $c_s$-cone of the low-energy fluctuations generally does not coincide with the UV light-cone), and ultimately the analytic structure of the high-energy amplitude should be added to the list of assumptions made about the UV completion (in addition to unitarity and Froissart-boundedness). A violation of \eqref{eqn:pos_LI} signals that the UV completion violates unitarity, locality (Froissart-boundedness) or causality, whereas a violation of \eqref{eqn:dt_pos1} or \eqref{eqn:dt_pos2} signals that the UV completion violates unitarity, locality or our assumption of analyticity. These assumptions are discussed in more detail in Appendix~\ref{app:derivation}.}

Finally, note that \eqref{eqn:dt_pos2} is not a strict inequality, since a low-energy amplitude may have $d_{st\omega\omega} = 0$ and still satisfy \eqref{eqn:posNLO_csst2}. 
In fact this is precisely what happens for the purely quartic Galileon ($c_3 = c_5 = 0$), whose tree-level amplitude is simply $Z^2 \mathcal{A} = c_2 c_4 ( s^3 + t^3 + u^3)$ on \emph{any} galileid background. 
Ordinarily, since both $\tilde{d}_3$ and $\tilde{d}_4$ change sign when $c_s$ crosses $1$, any region of parameter space which satisfies \eqref{eqn:dt_pos2} for subluminal backgrounds (i.e. has $d_{st\omega\omega} > 0$ on backgrounds with $c_s < 1$) will necessarily violate it on superluminal backgrounds (i.e. will have $d_{s t \omega \omega} < 0$ on backgrounds with $c_s > 1$), and vice versa. 
Only in the special case $d_{st\omega\omega} = 0$ can \eqref{eqn:dt_pos2} be satisfied by both sub- and super-luminal solutions in this covariant Galileon theory.
Put another way, if one starts from a unitary, causal, local, Lorentz invariant UV completion (with invariant speed $c=1$), then the EFT for low-energy fluctuations about a boost-breaking background will typically only be viable for a particular sign of $1 - c_s^2$, unless there is some cancellation which produces $d_{st\omega\omega} = 0$ in the amplitude.\footnote{Note that, throughout this paragraph, we have implicitly kept the $c_i$ fixed, while exploring different vacua by changing $\{\alpha,\beta\}$, subject to the tadpole and stability conditions being satisfied.}

\paragraph{The space of allowed vacua:}
\begin{figure}[t!]
\begin{center}
\includegraphics[width=.32\linewidth]{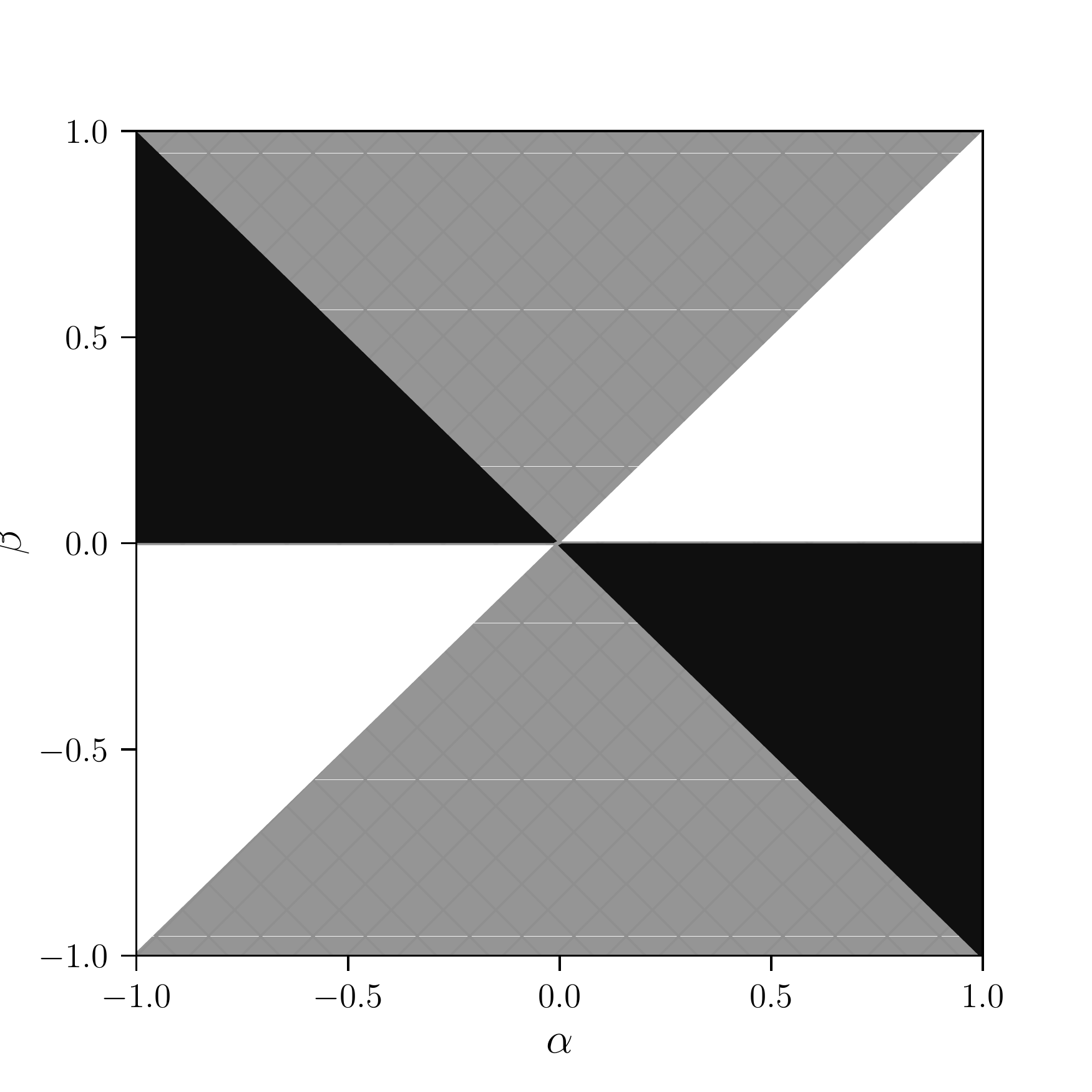}
\includegraphics[width=.32\linewidth]{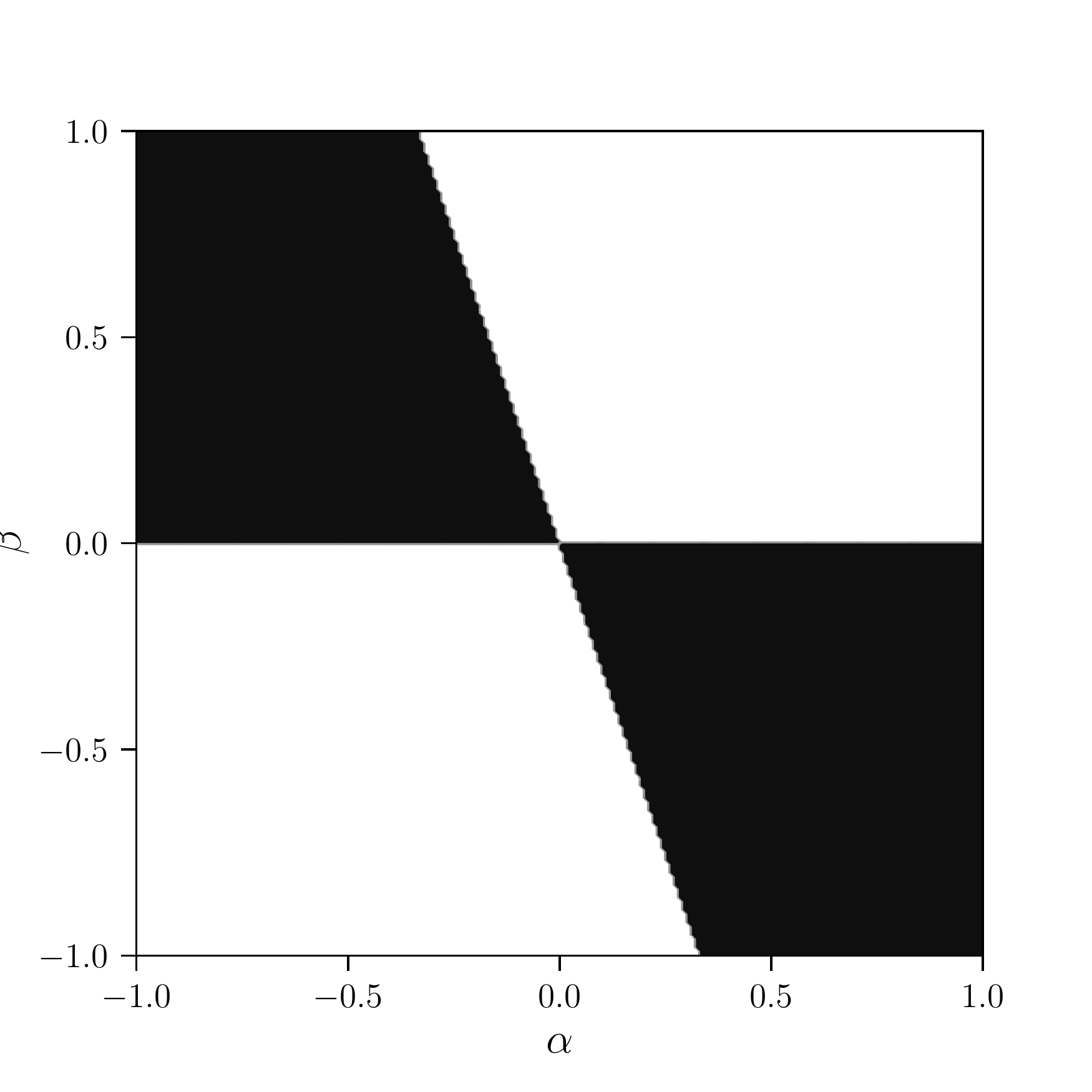}
\includegraphics[width=.32\linewidth]{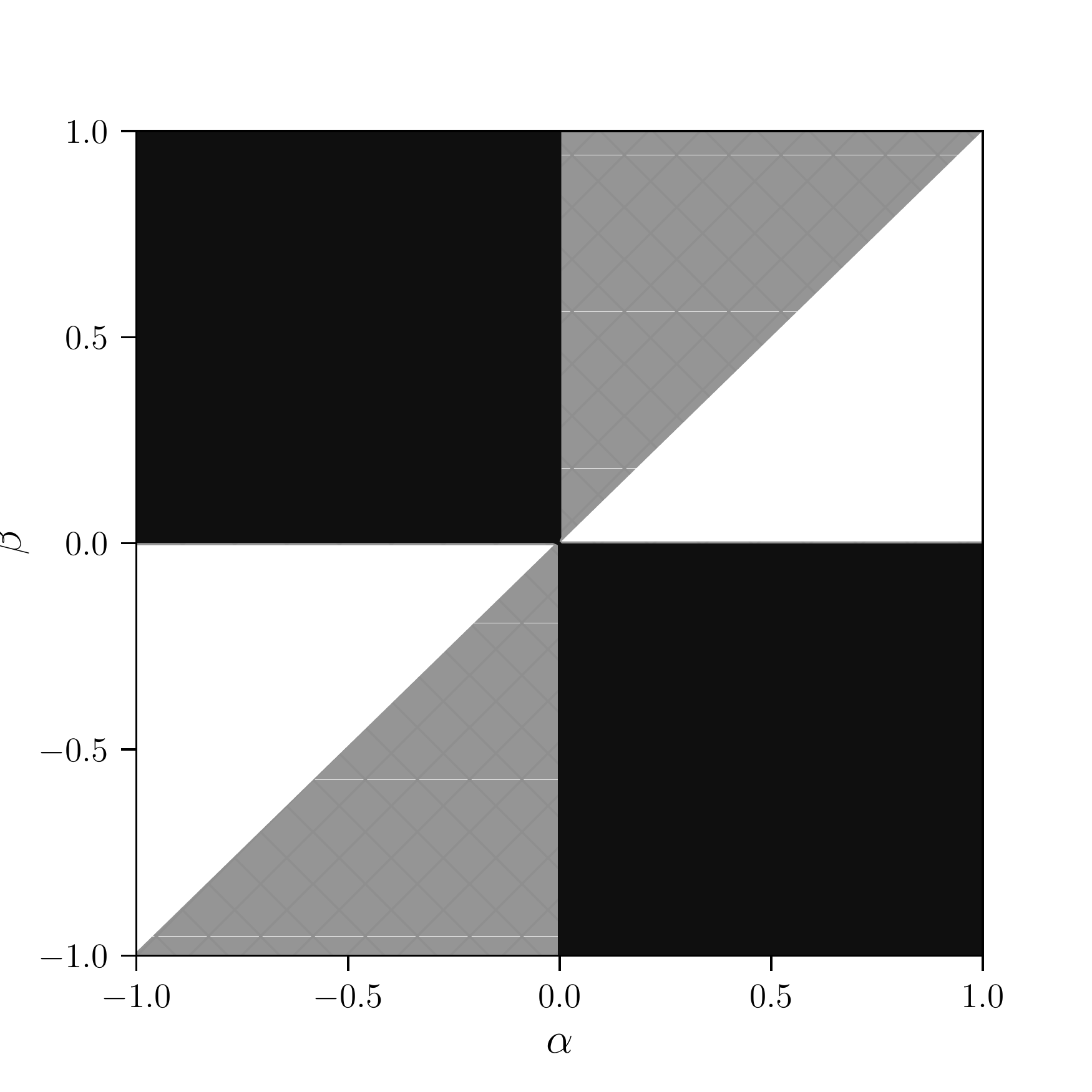}
\end{center}
\caption{
Here we show the space of vacuum solutions, as parametrised by the $\alpha$ and $\beta$ parameters. The three panels from left to right correspond to the cubic, quartic and quintic Galileon theories, respectively, where only the respective $c_i$ and $c_2$ are non-zero and the kinetic term has been normalised by setting $c_2 = -1$ (in anticipation of the next section and hence ruling out a stable vacuum solution $\bar \phi = 0$). For any given choice of $\left(\alpha,\beta\right)$ we solve the tadpole condition \eqref{Jcs} to obtain the free $c_i$. Regions ruled out by ghost and/or gradient instabilities are then shaded black, stable regions that fail either of the positivity bounds (\eqref{eqn:dt_pos1} and \eqref{eqn:dt_pos2}) are hatched grey and regions free of ghost and gradient instabilities and consistent with the positivity bounds considered here are white.
\label{fig-regionPlot-noCosmo}}
\end{figure}
Having derived positivity bounds and stability conditions for the family of Galileon vacuum solutions \eqref{eqn:galileid} above, we now investigate how restrictive and informative these criteria are in identifying physically well-motivated vacua. 
We first focus on a particular vacuum (with fixed $\alpha, \beta$) and use these bounds to constrain the remaining $c_n$, i.e. ask whether this particular solution could ever arise from any theory of the form \eqref{CovGal_action}.
In figure \ref{fig-regionPlot-noCosmo} we show the results for a cubic, quartic and quintic Galileon (panels from left to right), respectively. Having normalised the kinetic $\mathcal{L}_2$ term, these theories each have one free $c_i$ parameter, which we fix using the tadpole condition \eqref{Jcs}. We then consider the space of vacuum solutions \eqref{eqn:galileid} as parametrised by $(\alpha,\beta)$ and identify, which regions of parameter space are ruled out by ghost and gradient stability constraints (black regions in figure \ref{fig-regionPlot-noCosmo}). For the residual parameter space we then ask, whether the positivity bounds \eqref{eqn:dt_pos1} and \eqref{eqn:dt_pos2} are satisfied---regions failing this test (hatched gray regions) can be as large as $2/3$ of the residual parameter space, so positivity bounds are a powerful tool to identify physically well-motivated vacua (white regions) in this context. 

To develop intuition for the vacuum structure outlined in figure \ref{fig-regionPlot-noCosmo}, it is useful to discuss one example in more detail. We pick the cubic Galileon, i.e. the leftmost panel in figure \ref{fig-regionPlot-noCosmo}, and normalise the kinetic term by setting $c_2 = -1$ (in anticipation of the following section). From \eqref{Jcs} we obtain the tadpole condition for the relevant cubic Galileon vacua 
\begin{align} \label{tadpoleSol}
12 \beta \left(\alpha + \beta\right) c_3 = \alpha + 3 \beta,
\end{align}
which we can solve for $c_3$. Using this solution, from \eqref{Jcs} we can then find the following conditions for the absence of ghost and gradient instabilities, respectively
\begin{align}
\beta \left(\alpha + \beta\right) &> 0, 
&\alpha^2 + 2\alpha\beta + 3\beta^2 &\geq 0.
\end{align}
Only the no-ghost condition places a non-trivial constraint on the allowed vacua here, no real $\{\alpha,\beta\}$ violate the gradient stability condition.
Using \eqref{tadpoleSol} and imposing the above stability conditions, the two positivity bounds \eqref{eqn:dt_pos1} and \eqref{eqn:dt_pos2}, respectively, then imply
\comment{
\begin{align}
&\beta^2 \left(\alpha + \beta\right)^2 \left(\alpha + 3 \beta\right)^2 \left(\alpha^2 + 2 \alpha \beta + 3 \beta^2\right) \left(2 \alpha^2 + 4 \alpha \beta + 3 \beta^2\right) > 0
 \nn \\
&\frac{\left(\alpha - \beta\right) \left(\alpha + 3 \beta\right)}{\beta^2 \left(\alpha + \beta\right)^2 \left(\alpha^2 + 2 \alpha \beta + 3 \beta^2\right)} \geq 0.
\end{align}
}
\begin{align}
2 \alpha^2 + 4 \alpha \beta + 3 \beta^2 &> 0,
&\left(\alpha - \beta\right) \left(\alpha + 3 \beta\right) &\geq 0.
\label{example_posBounds}
\end{align}
The first positivity condition is satisfied for all stable vacua, so places no further constraints here. The second positivity bound, on the other hand, does place a significant constraint. The boundaries of the `positive' region in the cubic Galileon panel of Fig. \ref{fig-regionPlot-noCosmo} are ultimately determined by the factor of $(\alpha - \beta)$ in the second positivity bound and $\beta$ in the no-ghost stability bound ---the boundaries are correspondingly the lines $\alpha = \beta$ and $\beta = 0$. 
When viewed in this $\{\alpha,\beta\}$ parameter space, positivity bounds therefore eliminate 2/3 of the parameter space consistent with stability bounds alone for this example. 

Instead of scanning vacua in the $\{\alpha,\beta\}$ space and asking whether they could arise from (a subset of) covariant Galileon theories, we can instead consider a  particular covariant theory \eqref{CovGal_action} (with fixed coefficients $c_n$) and 
use positivity bounds to effectively constrain the allowed range of $\beta$, i.e. probe which vacua in this particular EFT could be compatible with a standard UV completion.
Focusing on the second positivity bound \eqref{eqn:dt_pos2}, when written in terms of the original $c_n$ appearing in the covariant action \eqref{CovGal_action}, this bound implies that,
\begin{align}
(1 - c_s^2 ) \, \frac{  8 \beta^3 c_3^3 + 12 \beta c_3 \, \beta^2 c_4 - \beta^3 c_5  }{ \beta c_3 + 6 \beta^2 c_4 - 2 \beta^3 c_5 } \leq 0  \; .
\label{eqn:egpos2}
\end{align}
for any Galileid vacua in which now $\alpha$ is fixed by the tadpole condition ($J=0$) and $\beta$ obeys the no-ghost stability condition ($Z^2/c_s^3 > 0$, cf. \eqref{Jcs}),
\begin{align}
c_2 + 12 \beta c_3 + 36 \beta^2 c_4 - 8 \beta^3 c_5 > 0 \; . 
\label{eqn:egZ2}
\end{align} 
Note that there is the freedom to redefine $\phi \to C \phi$ for any real constant $C$ (which is equivalent to mapping $\beta \to C \beta$ and $c_n \to C^n c_n$) and indeed physical quantities like $c_s^2$ and the bounds \eqref{eqn:egpos2} and \eqref{eqn:egZ2} are not affected by such a rescaling. 
A crucial observation is that, for both positivity bounds (including the first bound not explicitly shown in the $c_i$ basis above) and once $\alpha$ has been solved for using the tadpole condition, all $c_i$ always enter in the combination $\beta^{i-2} c_i$. By construction, the same is true for ghost and gradient stability bounds, so when written in this basis, the bounds for a theory with $n$ non-zero $c_i$ only depend on $n$ independent   effective parameters. This is different to the $\{\alpha,\beta\}$ space considered above, where an additional (dependent) parameter was present.

We now again consider the simplest models in which only a single Galileon interaction is present. For example, if $c_4 = c_5 = 0$, so only the cubic interaction is present, then all $12 \beta c_3 > -c_2$ vacua obey \eqref{eqn:egZ2} and are stable, but only the superluminal ones ($1 - c_s^2 \leq 0$) satisfy \eqref{eqn:egpos2}. Which vacua are sub/super-luminal depends on the sign of $c_2$. 
When $c_2 > 0$ the stable superluminal vacua correspond to $-c_2 <  12\beta c_3 \leq 0$, and when $c_2 < 0$ the stable superluminal vacua correspond to $|c_2| < 12 \beta c_3 \leq 2 |c_2|$. 
In both cases, the positivity bound \eqref{eqn:egpos2} has greatly reduced the available parameter space compared to stability considerations alone (from a semi-infinite range to a finite line segment\footnote{
For theories in which $c_2 = 0$, so the kinetic term for fluctuations comes entirely from the cubic Galileon interaction, $c_s$ is always subluminal and so the positivity bounds can never be satisfied---in this case the line segment shrinks to nothing. 
}). 
Similarly, for the simple model in which only the quintic Galileon is present ($c_3 = c_4 = 0$), all $8 \beta^3 c_5 < c_2$ vacua obey \eqref{eqn:egZ2} and are stable, but again only the superluminal ones satisfy the positivity condition \eqref{eqn:egpos2}. 
When $c_2 > 0$ the stable superluminal vacua are $0 < 8\beta^3 c_5 < c_2$, and when $c_2 < 0$ the stable superluminal vacua are $- 4 |c_2| < 8 \beta^3 c_5 < -|c_2|$.
So again positivity has reduced the available parameter space to a finite window. 
Finally, we already noted above that the quartic Galileon is a special case, since the second positivity bound is trivially saturated there (giving $0 \geq 0$). The first positivity bound, on the other hand, is always satisfied by any solution that respects the  no-ghost stability condition, $\beta^2 c_4 > -c_2/36$. In this special case, both sub- and super-luminal solutions are consistent with  all bounds.
%

\section{Interplay with cosmological observations}
\label{sec:3}

In this section we explore the interplay between positivity bounds and observational constraints for Galileon theories of cosmological interest. To that end, we will first discuss the observational bounds by themselves and then ask how informative the above novel positivity bounds (computed around a specific family of boost-breaking vacua) can be for cosmologically relevant dark energy theories theories (around a different boost breaking vacuum, i.e. the FLRW solution).

\subsection{Observational constraints for the Galileon}
\label{sec:observations}

\paragraph{Cosmological background solutions:} 
When considering cosmological background solutions for the Covariant Galileon, the existence of a so-called `tracker' is crucial \cite{DeFelice:2010pv}. Here we briefly summarise its characteristics and consequences as discussed in detail in \cite{DeFelice:2010pv,Barreira:2013jma,Renk:2017rzu}. The tracker is characterised by the constant $\xi$ parameter, where one can define
\begin{align} \label{xiDef}
\xi \equiv \frac{\dot\phi H}{\Lambda_3^3} = \text{constant}.
\end{align}
Throughout this section, $\Lambda_3 = ( M_P H_0 )^{1/3}$ where $H_0$ is the Hubble rate today.
The existence of this solution is tightly linked to the underlying shift symmetry of the theory\footnote{Note this is simply a consequence of the $\phi \to \phi + c$ shift symmetry part of the full Galilean symmetry, i.e. the full Galilean symmetry (weakly broken by the coupling to matter and gravity) is not relevant here.}, which allows re-writing the field equations as a current conservation equation. More specifically, one finds
\begin{align} \label{currentCons}
\nabla_\mu {\cal J}^\mu &= \dot{\cal J}^0 + 3 H {\cal J}_0 = 0,
\\ \label{current}
{\cal J}_0 &= c_2 \xi - 6 c_3 \xi^2 + 18 c_4 \xi^3 + 5 c_5^{\text{}} \xi^4.
\end{align}
As a consequence of \eqref{currentCons}, ${\cal J}^0$ decays as $a^{-3}$ and so there exists a `tracker' solution.
In \cite{Barreira:2013jma} it was shown that reaching this tracker solution while the fractional contribution of dark energy to the energy density of the Universe, $\Omega_{\rm DE}$, is still sub-dominant is required in order to comply with CMB constraints. In what follows we will therefore assume this tracker has indeed been reached sufficiently early, effectively setting ${\cal J}^0 = 0$ and using \eqref{current} as a constraint on the $c_i$. Assuming a spatially flat Universe, one can furthermore use the Friedmann equation to derive
\begin{align} \label{OmegaPhi}
\Omega_{\phi,0} = \frac{1}{6} c_2^{\text{}} \xi^2 - 2 c_3^{\text{}} \xi^3 + \frac{15}{2} c_4^{\text{}} \xi^4 + \frac{7}{3} c_5^{\text{}} \xi^5.
\end{align}
Note that we are considering self-accelerating solutions here, where the Galileon scalar $\phi$ makes up the entirety of dark energy, i.e. $\Omega_{\rm DE} = \Omega_{\phi}$. The $0$ index refers to the time $t_0$, i.e. to today.
The relations \eqref{current} and \eqref{OmegaPhi} then have profound consequences for the background evolution of the Covariant Galileon. In essence, they will tightly constrain one of the free coefficients of the theory, as we can use \eqref{current} to solve for $\xi$ or one of the free $c_i$ and then use \eqref{OmegaPhi} to place a tight observational constraint on (another) one of the $c_i$ for a given set of bounds on $\Omega_{\phi,0}$. 
Specifically, we can use that $\Omega_{\phi,0} \sim 0.7$ \cite{Planck:2018vyg}.

\paragraph{Cosmological backgrounds and (un-)stable vacua:}
The tracker constraints discussed above have immediate consequences for vacuum stability, i.e. which solutions satisfy the ghost and gradient stability conditions derived from \eqref{Jcs}. As foreshadowed in the previous section, perhaps the most interesting consequence relates to the instability of the trivial vacuum solution, $\bar \phi = 0$. Here it is instructive to start with the same simple examples used to illustrate the `space of vacua' in the previous section: pure cubic, quartic and quintic Galileons, each case with only two free parameters: $c_2$ and the respective $c_i$. 
Using \eqref{OmegaPhi} together with ${\cal J}^0 = 0$ and \eqref{current}, we find
\comment{
\begin{align}
&\text{cubic Gal.}: &\Omega_{\phi,0} &= -\frac{c_2}{6} \xi^2 \nn \\
&\text{quartic Gal.}: &\Omega_{\phi,0} &= -\frac{c_2}{4} \xi^2 \nn \\
&\text{quintic Gal.}: &\Omega_{\phi,0} &= -\frac{3c_2}{10} \xi^2.
\end{align}
}
\begin{align}
\Omega_{\phi,0}^{\rm cubic} &= -\frac{c_2}{6} \xi^2,
&\Omega_{\phi,0}^{\rm quartic} &= -\frac{c_2}{4} \xi^2, 
&\Omega_{\phi,0}^{\rm quintic} &= -\frac{3c_2}{10} \xi^2.
\end{align}
Since $\xi$ is a real constant \eqref{xiDef}, obtaining a positive $\Omega_{\phi, 0}$ therefore requires a negative $c_2$ for all of these simple examples. When considering the full space of Galileon theories with arbitrary $c_i$ combinations, this conclusion no longer follows just from the tracker conditions, but overall observational constraints still enforce a negative $c_2$ \cite{Barreira:2013jma}.  
Crucially this disconnects cosmologically relevant Galileons from the trivial $\bar \phi  = 0$ vacuum. In our present positivity-related context, this means that bounds computed around a $\bar \phi  = 0$ background do not apply to the cosmologically relevant branch of solutions, further motivating our present exploration of bounds for boost-breaking vacua. 

\paragraph{Data sets and priors:} We now perform a Markov chain Monte Carlo (MCMC) analysis, computing cosmological constraints on the Galileon model parameters. The data sets we use come in two parts: First, CMB temperature and polarisation data from Planck 2018 \cite{Planck:2019nip}. More specifically, we use the high-$\ell$ TTTEEE, low-$\ell$ EE, and low-$\ell$ TT likelihoods. Note that we therefore do not use the pre-marginalised {\it Plik\_lite} likelihood, which has mostly been shown to be accurate for $\Lambda{}$CDM cosmologies (i.e. not for the Galileon cosmologies considered here). However, in practice we find that the resulting parameter constraints we will discuss below are almost unaffected by the choice between the pre-marginalised {\it Plik\_lite} and `full' Planck likelihoods. 
We complement the above CMB data by using BAO measurements from the 6dF Galaxy Survey \cite{Beutler:2011hx}, SDSS DR7 LRG \cite{Padmanabhan:2012hf} and from BOSS DR9 CMASS \cite{Anderson:2012sa}. Note that we have not included newer, additional BAO data sets such as from the SDSS main galaxy sample \cite{Ross:2014qpa}, which are in mild $\sim 2\sigma$ tension with the Galileon cosmologies discussed here---see \cite{Renk:2017rzu} for a detailed discussion of this choice of BAO datasets. Modulo using current Planck data, we are therefore using the same data sets as \cite{Renk:2017rzu} here.
With this minimal set of cosmological data (we will discuss the impact of including additional observations/data sets below), we then proceed as follows. We fix $c_1 = 0$ for simplicity and normalise $c_2 = -1$, where we recall a negative $c_2$ is required by cosmological constraints \cite{Barreira:2013jma}. We then use \eqref{current} and \eqref{OmegaPhi} to solve for $\{c_4,c_5\}$, so that $\{\xi,c_3\}$ are the residual free Galileon parameters. As part of the MCMC analysis we now vary these two parameters in addition to the standard  $\Lambda{\rm CDM}$ parameters $\{\Omega_{\rm b}, \Omega_{\rm cdm}, H_0, A_s, n_s, \tau_{\rm reio}\}$ and $\Sigma m_\nu$ (the sum of neutrino masses). Note that we assume a degenerate neutrino mass spectrum, where a (significantly) non-zero sum of neutrino masses is observationally mandated in Galileon cosmologies \cite{Barreira:2014jha}.
No prior boundaries are imposed on any of the varied parameters, except for requiring a non-negative sum of neutrino masses and $\tau_{\rm reio} \geq 0.004$ (corresponding to $z_{\rm reio} \gtrsim 6$ and motivated
by observations of the Gunn-Peterson trough, see e.g. \cite{SDSS:2001tew}).
\\

\begin{figure}[t!]
\begin{center}
\includegraphics[width=.49\linewidth]{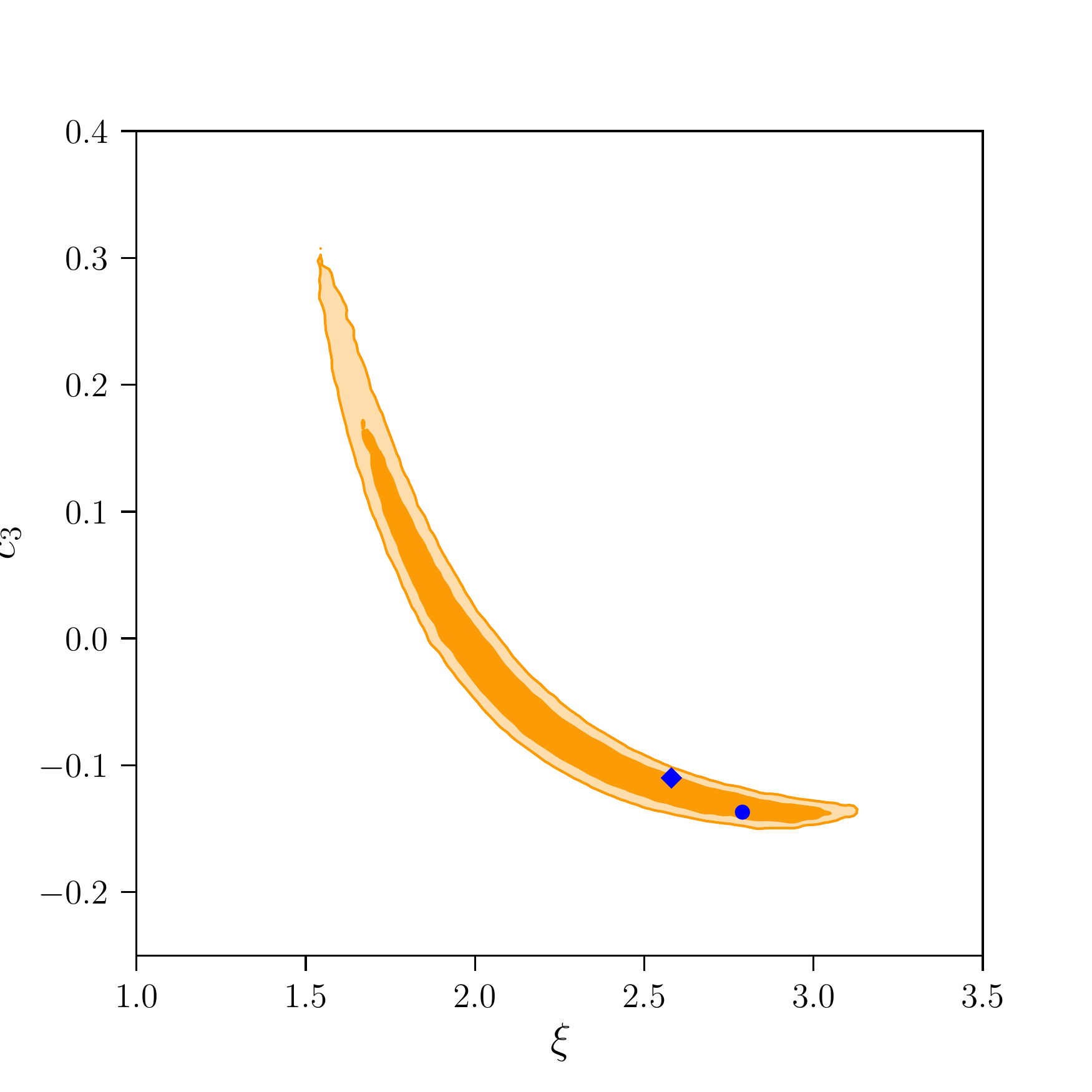}
\end{center}
\caption{
Here we show cosmological data constraints in the $\xi-c_3$ parameter space for covariant Galileons. We fix $c_2 = -1$ to remain consistent with observational constraints, while $c_4$ and $c_5$ are fixed by the current conservation \eqref{current} and Friedmann equations \eqref{OmegaPhi}. 
Data contours mark $68\%$ and $95\%$ confidence intervals, computed using CMB+BAO data (see main text for details). The blue dot and diamond identify the CMB + BAO and CMB + BAO + ISW best-fit cosmologies, respectively (where the best-fit cosmology including ISW constraints has been lifted from \cite{Renk:2017rzu}). 
\label{fig-pureDataPlot}}
\end{figure}
\paragraph{Cosmological parameter constraints:} 
Throughout this paper we will focus on constraints in the $\{\xi,c_3\}$ plane, marginalising results over $\{\Omega_{\rm b}, \Omega_{\rm cdm}, H_0, A_s, n_s, \tau_{\rm reio}, \Sigma m_\nu\}$. Working with the full covariant Galileon and solving the current conservation \eqref{current} and Friedmann equations \eqref{OmegaPhi} for $c_4$ and $c_5$, we are then left with $\xi$ and $c_3$ as the two parameters controlling dark energy dynamics here (recall that we have fixed $c_1 = 0$ for simplicity and normalised $c_2 = -1$ as required by cosmological constraints \cite{Barreira:2013jma}). In figure \ref{fig-pureDataPlot} we show the resulting parameter constraints in the $\xi{}-c_3$ plane, which are in excellent agreement with the corresponding results found by \cite{Renk:2017rzu}, as expected. As a reference point for later, we are also marking the best-fit parameter values to the data for the CMB+BAO constraints considered here (circular dot) as well as for the CMB+BAO+ISW constraints from \cite{Renk:2017rzu} (diamond). However, we caution against overinterpreting these best-fit points and emphasise that any region of parameter space within the $\sim 2 \sigma$ contours in figure \ref{fig-pureDataPlot} should be considered as fully consistent with the observational bounds considered here.
Overall CMB+BAO constraints require $1.5 \lesssim \xi \lesssim 3$ and $-0.2 \lesssim c_3 \lesssim 0.3$ at the $2\sigma$ level.\footnote{Note that the Einstein-Boltzmann solver we use, hi\_class \cite{Zumalacarregui:2016pph,Bellini:2019syt}, struggles to solve cosmologies with $|c_3| \lesssim 0.005$ in the setup discussed here. This is because the scalar equation of motion and the `spatial' Friedman equation are used to solve for $\ddot\phi$ and $\dot H$ in the default code setup, but this solution becomes degenerate at a point during the cosmological evolution for small values of $c_3$ (and when solving for $c_4, c_5$ as discussed above). However, this is an artefact of the solver and not related to any underlying physical singularity (one could e.g. use the other Friedman equation to solve for $H$ instead), so we here simply interpolate across the $|c_3| \lesssim 0.005$ region.}
\\

\subsection{Positivity bounds as theoretical priors}
\label{sec:priors}

After the observational considerations above, we are now in a position to discuss the interplay between observational bounds on the covariant Galileon as a cosmological candidate theory for dark energy and the positivity bounds derived in the previous section. 

\begin{figure}[t!]
\begin{center}
\includegraphics[width=.32\linewidth]{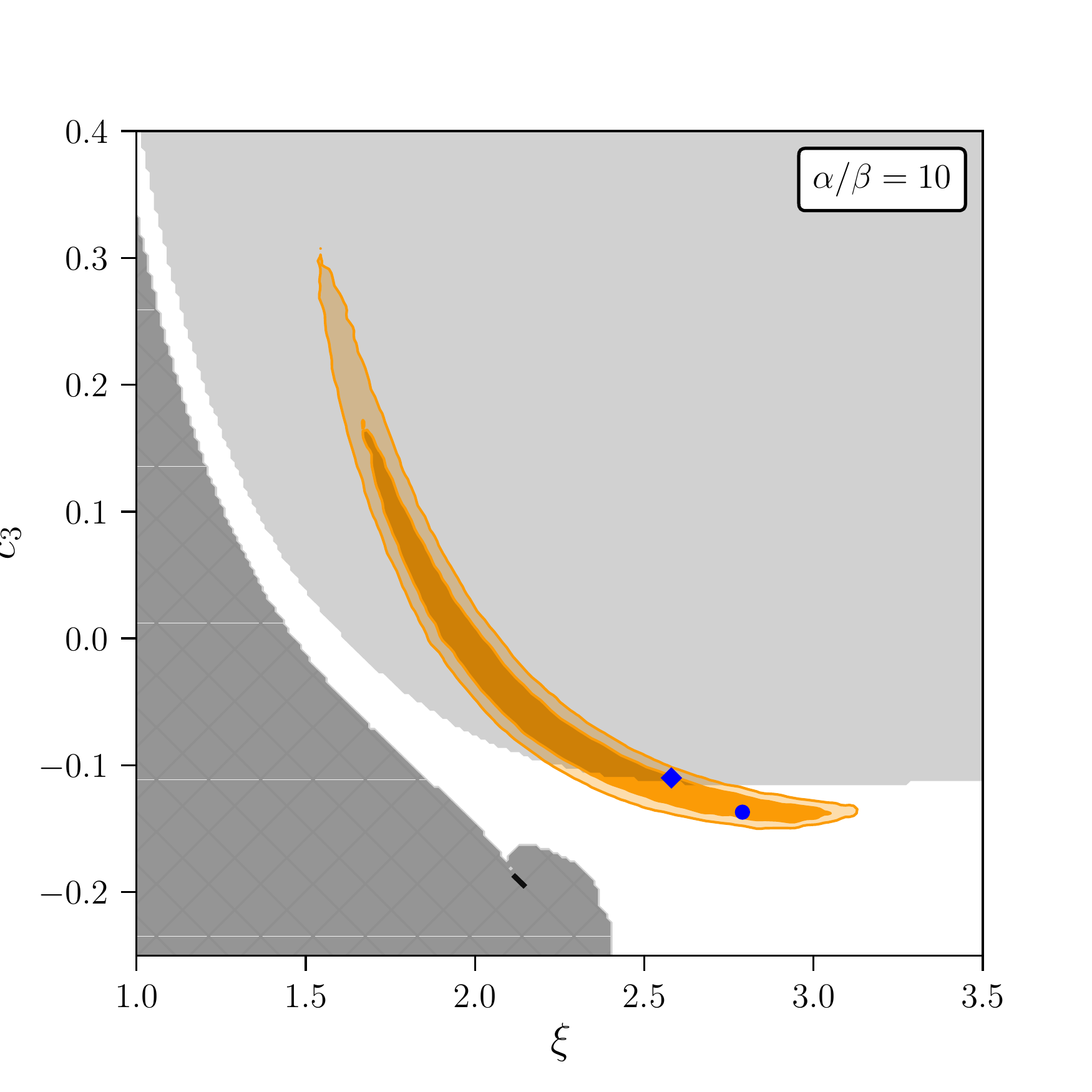}
\includegraphics[width=.32\linewidth]{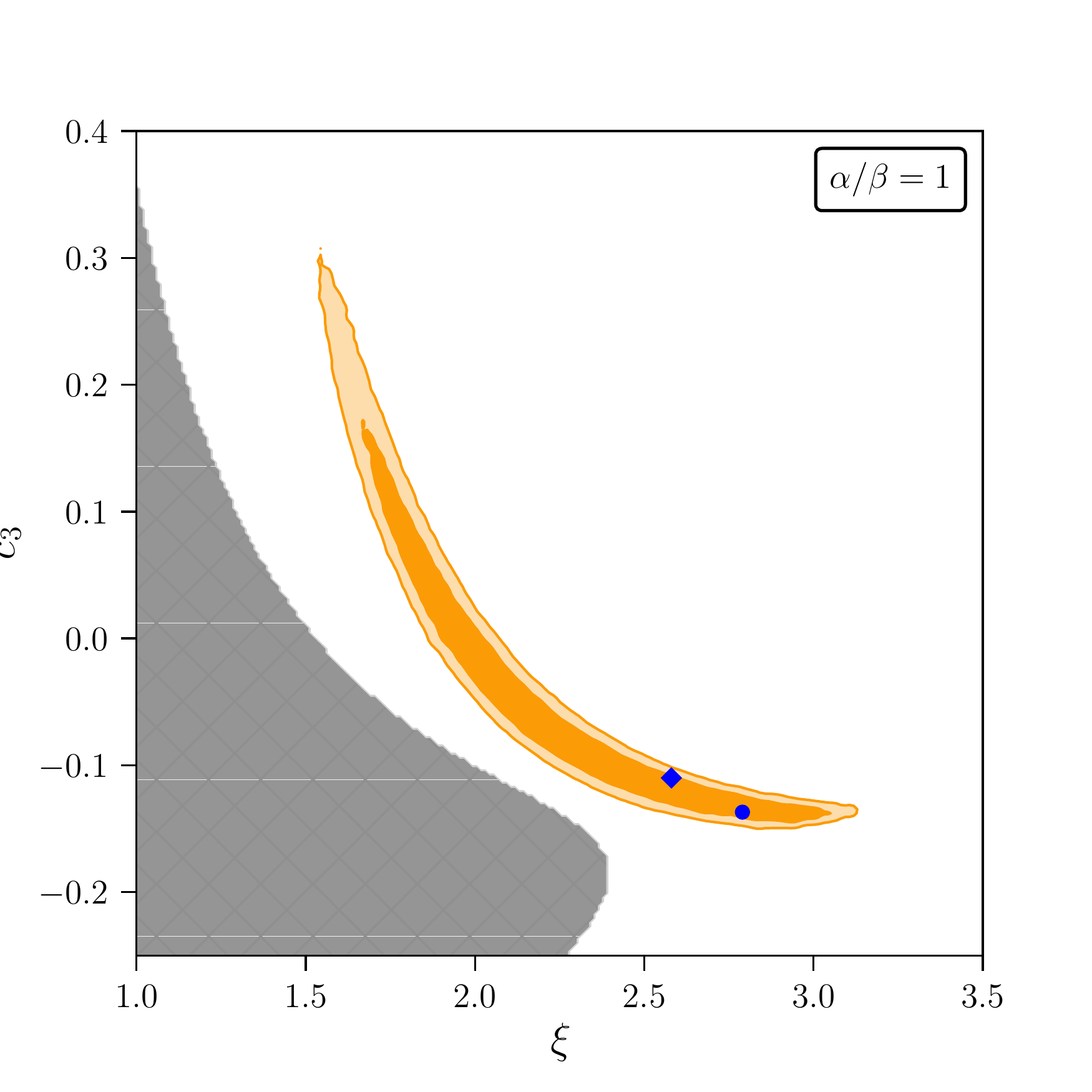}
\includegraphics[width=.32\linewidth]{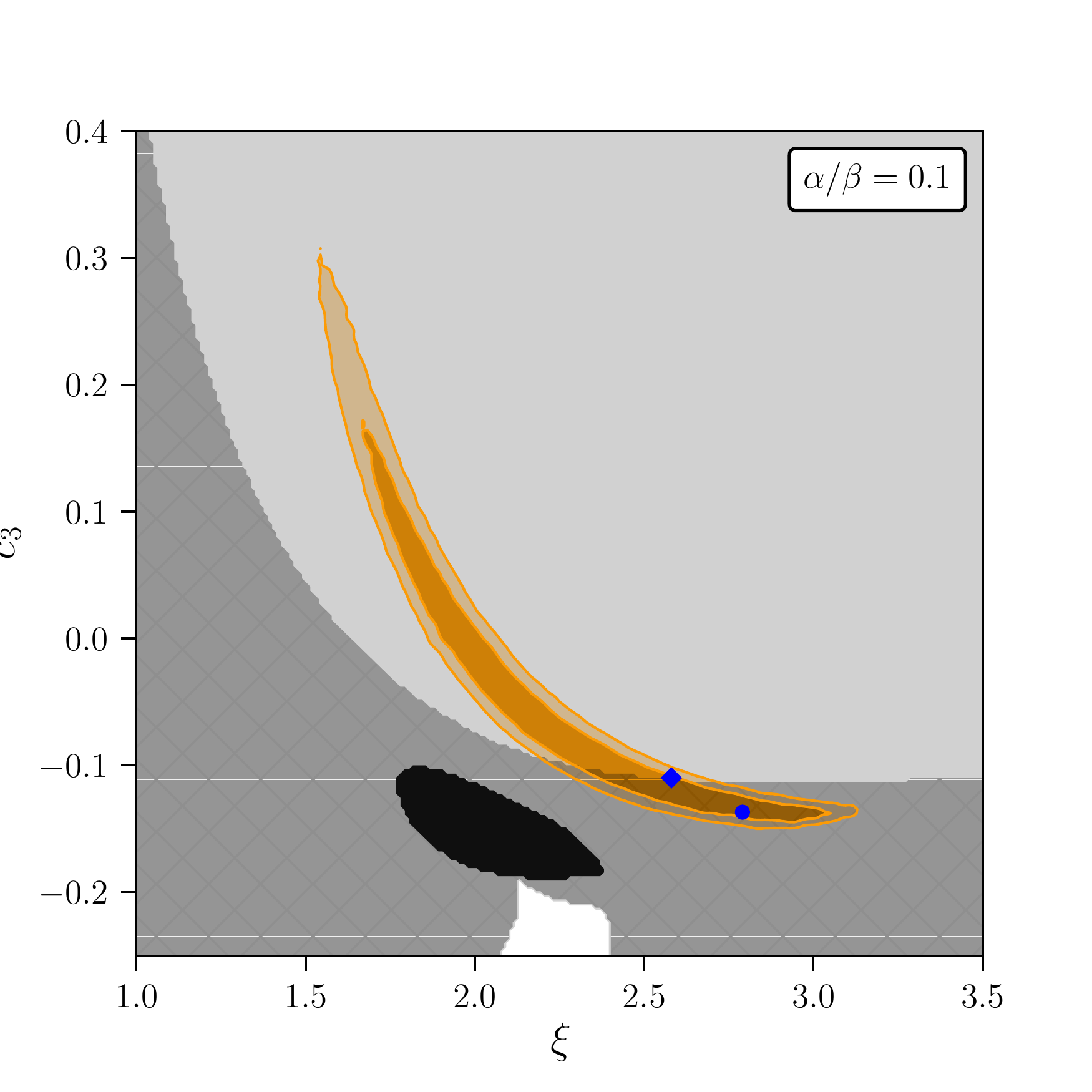}\\
\includegraphics[width=.32\linewidth]{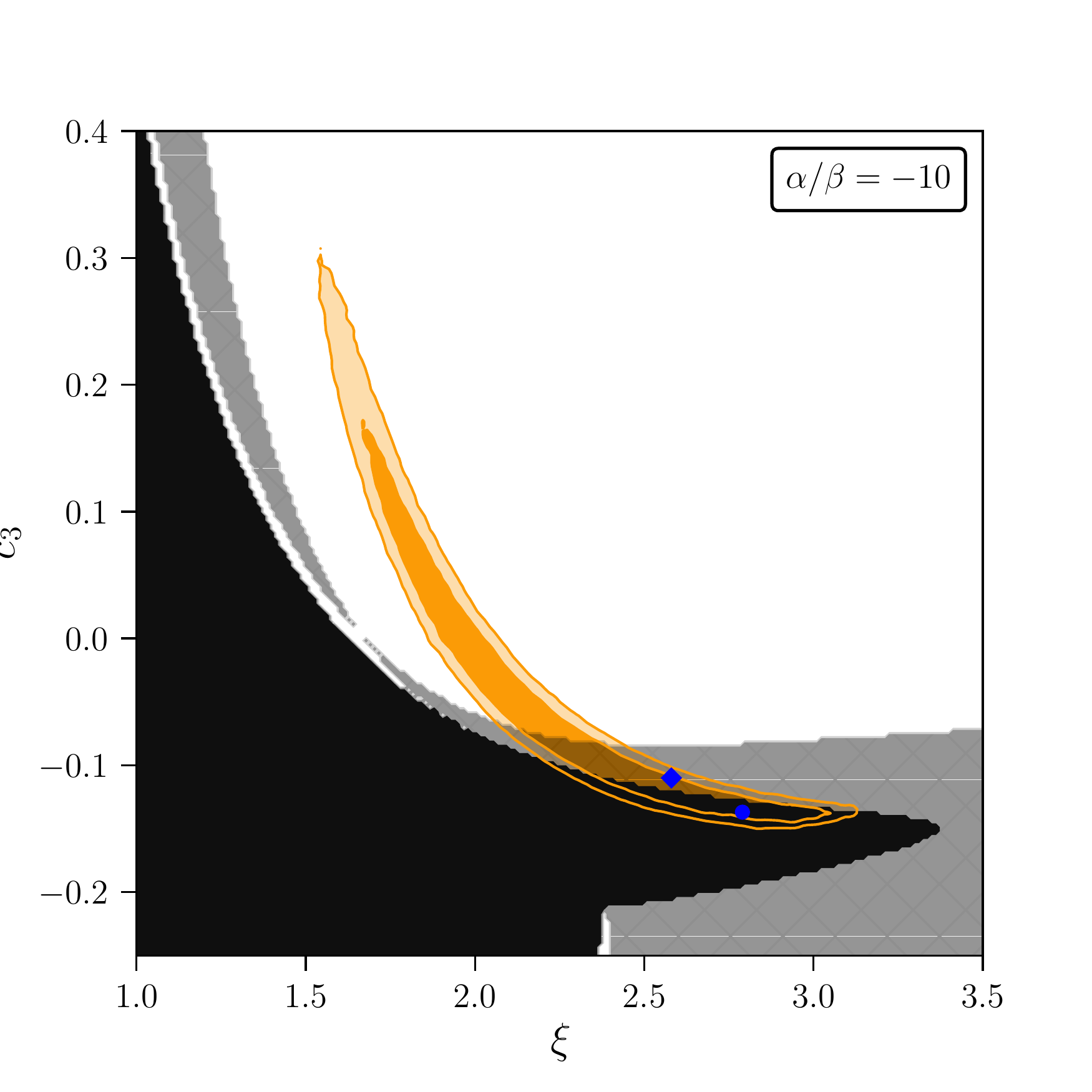}
\includegraphics[width=.32\linewidth]{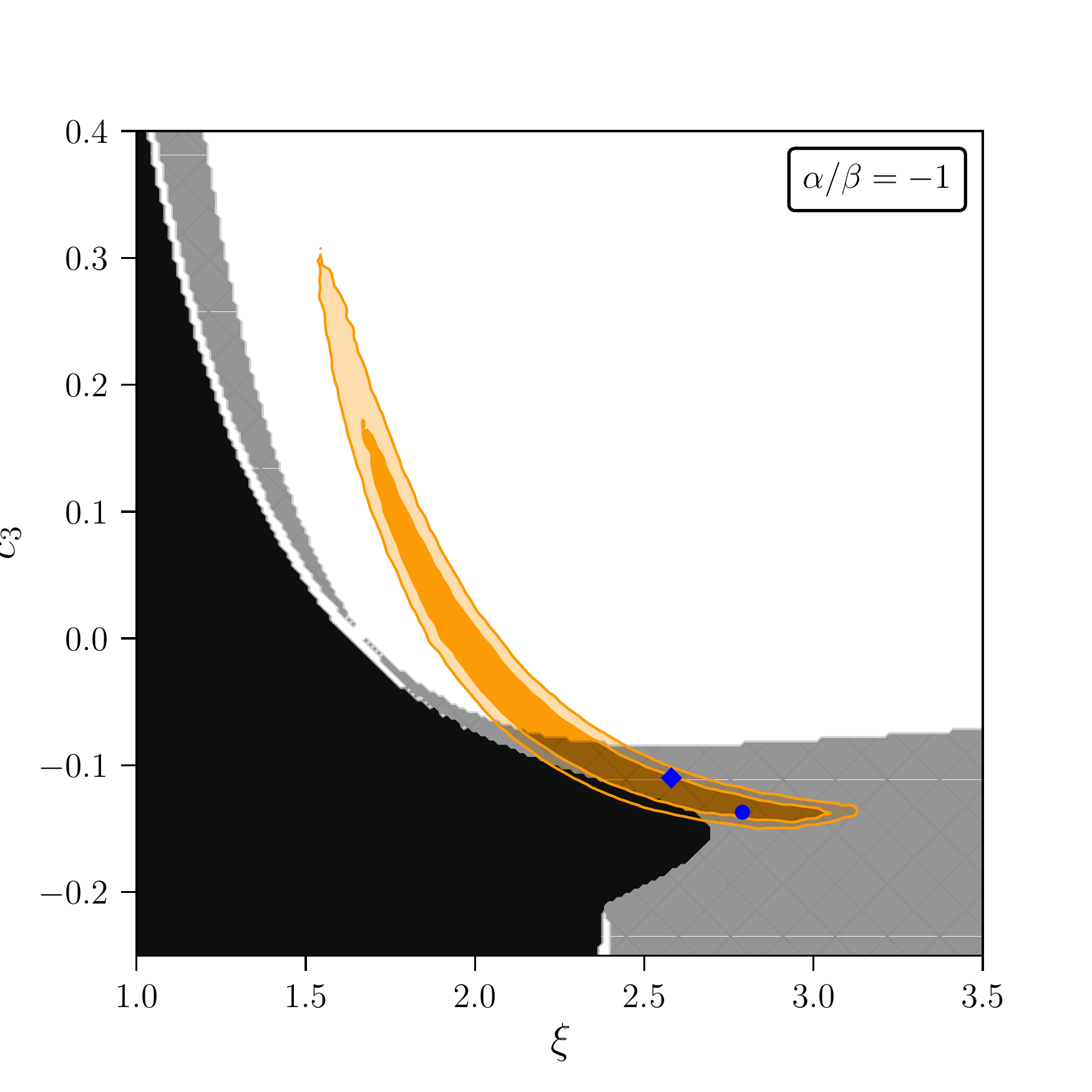}
\includegraphics[width=.32\linewidth]{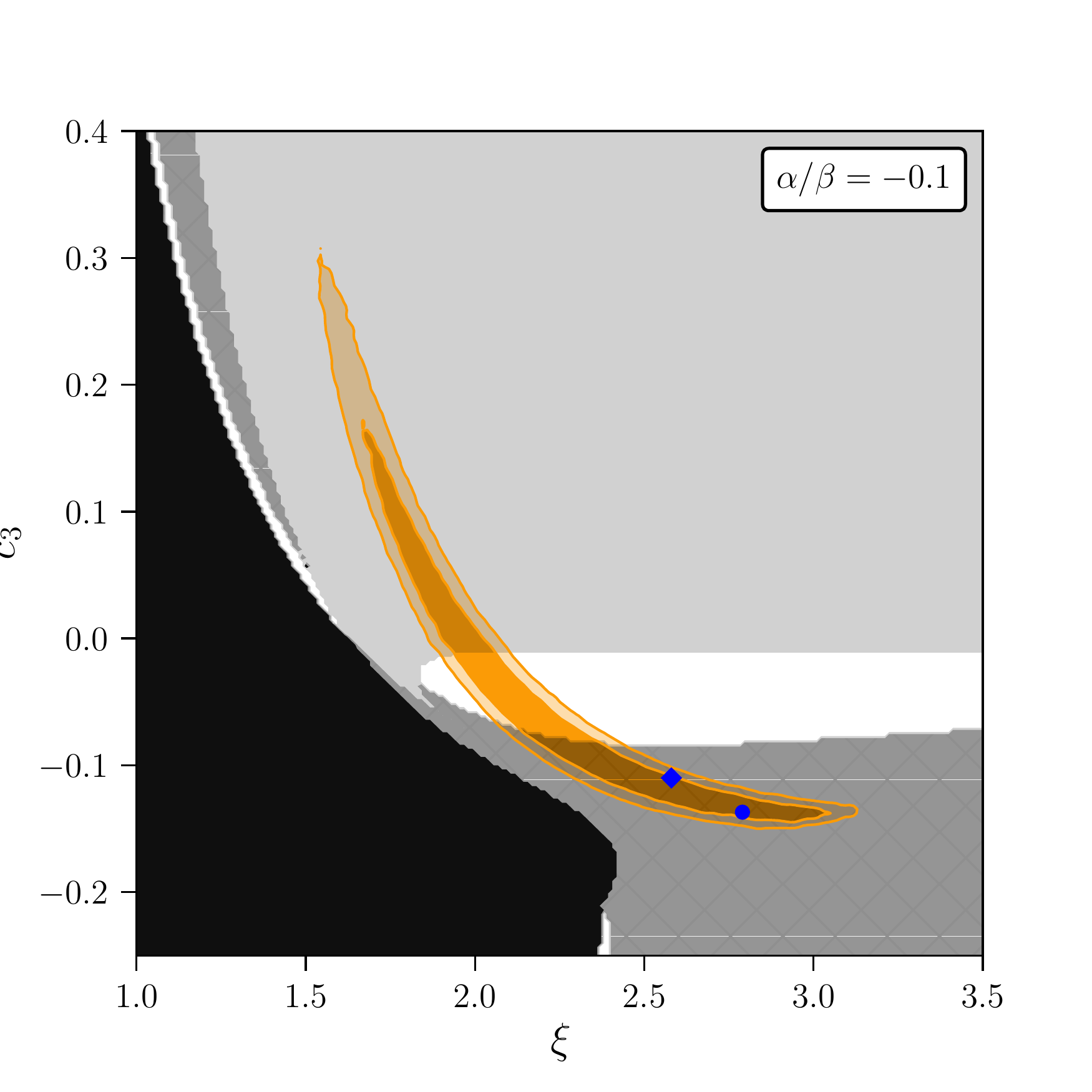}
\end{center}
\caption{
Here we illustrate the interplay between positivity bounds and cosmological stability constraints on the observationally relevant $\xi-c_3$ parameter space for covariant Galileons. 
Data contours are as in figure \ref{fig-pureDataPlot} and the different panels correspond to different families of vacuum solutions. {\bf Top row} (from left to right): $\alpha/\beta = 10, 1, 0.1$, {\bf Bottom row}: $\alpha/\beta = -10, -1, -0.1$. The middle panel in the top row therefore shows boost-invariant solutions, while all the other panels display different cross-sections of boost-breaking solutions.
Regions in black are those for which no stable background solution exists, i.e. the ghost or gradient constraint \eqref{Jcs} is violated. Dark grey, hatched regions are those, where one or more stable background/vacuum solutions exist, but they all fail their respective positivity bounds. Light grey then denotes regions of parameter space, where again stable vacua exist and some (but not all) of them satisfy positivity bounds. Finally, white regions are those, where stable vacuum solutions exist and all stable solutions satisfy their respective positivity bounds. 
\label{fig-regionPlot-boostBreak}}
\end{figure}

\paragraph{Positivity bounds and the space of vacua:}
Following the spirit of \cite{Melville:2019wyy}, we would like to know whether positivity bounds can be folded into a cosmological constraint analysis as informative theoretical priors. In addition we here specifically explore what this can tell us about the space of galileid vacua consistent with both observational constraints and positivity bounds. In order to explore this space of vacuum solutions, we would like to identify galileid vacua that can be associated to each point in the observationally relevant $\{\xi,c_3\}$ plane and we therefore proceed as follows: 
\begin{enumerate}
\item We fix $c_1 = 0$ for simplicity and normalise $c_2 = -1$, where we recall a negative $c_2$ is required by cosmological constraints \cite{Barreira:2013jma}. This leaves us with three free $c_i$ ($c_3,c_4,c_5$) and two free parameters ($\alpha$ and $\beta$ from \eqref{eqn:galileid}) tracing over different vacuum solutions.
\item We now choose an array of 3-tuples $\left\{\xi,c_3,\alpha/\beta \right\}$ as input\footnote{It is worth noting, that we could of course have chosen a different set of three input parameters, e.g. $\left\{c_3,c_4,\beta\right\}$. In particular, we here chose $\xi$ as one of our `native' parameters, since we are primarily interested in observationally relevant regions of parameter space for which a real $\xi$ solution must exist. 
}, solving the current conservation \eqref{current} and Friedmann equations \eqref{OmegaPhi} for $c_4$ and $c_5$, where we impose $\Omega_\phi= 0.69$ consistent with current constraints \cite{Planck:2018vyg}.\footnote{Note that we have checked that the resulting `labels' in step 4 (and hence the corresponding positivity priors) only change minimally when altering the fiducial value of $\Omega_\phi$ within the $2\sigma$ bound from Planck on $\Omega_\Lambda$, namely $0.6847 \pm 0.0146$ \cite{Planck:2018vyg}.}
\item Taking those solutions, we now find all values for $\alpha$ consistent with this from \eqref{Jcs}, in other words we find all the vacua of the form \eqref{eqn:galileid} consistent with the above solution and input $\left\{\xi,c_3,\alpha/\beta\right\}$. Note that there are up to four real solutions for $\alpha$ and that the value of $\alpha$ together with the $\alpha/\beta$ input also specifies $\beta$ here, i.e. we have now found all the vacua consistent with the input.
\item All the above solutions and vacua are now labelled according to whether I) they respect ghost and gradient stability conditions from \eqref{Jcs}, II) whether the associated positivity bounds \eqref{eqn:dt_pos1} and \eqref{eqn:dt_pos2} are satisfied.
\end{enumerate}
Following the above algorithm, the corresponding results are shown in figure \ref{fig-regionPlot-boostBreak}. Here we combine an analysis of the `positivity' and stability properties of the $\left\{\xi,c_3,\alpha/\beta\right\}$ parameter space as outlined above with observational constraints on the covariant galileon as a self-accelerating dark energy field. 
\blue{We highlight that, in combining positivity and stability bounds from given $\{\alpha,\beta\}$ vacua with observational constraints for cosmological vacua, we are here implicitly {\it assuming} that these vacua, as identified in the above fashion, co-exist within the same over-arching EFT and explore what the resulting combined bounds (from both classes of vacua) would be for the cosmologically relevant parameter space -- we will discuss this assumption in more detail below.} 
For the cases considered here, we find that, when imposing positivity bounds to require a `standard' UV completion as described in the previous section, the large-$\xi$ part of the observationally acceptable region is in tension with positivity bounds for vacua with $\alpha/\beta \lesssim 0.1$. This tension will become especially relevant when we discuss which of these vacua are `most relevant' in a cosmological context below.  
This shows that requiring the co-existence of several classes of galileid vacua and cosmological galileon dark energy solutions can be used to place strong constraints on cosmological parameter spaces. Alternatively, reversing the direction of the argument, observational constraints can be used in this way to place strong contraints on the space of physically acceptable vacua.
~\\

\paragraph{Identifying cosmologically relevant vacua and positivity bounds:}
Above we were considering positivity bounds derived from generic galileid vacua classified by their value of $\alpha/\beta$, investigating the interplay of these bounds with cosmological data constraints for galileon dark energy. If vacua with a given value of $\alpha/\beta$ exist within the same EFT as the cosmological background solution, we can then use the corresponding positivity bounds as informative theoretical priors for cosmological parameter constraints (if we have good reason to demand the existence of the relevant galileid vacua) or can use constraints from cosmological data to identfy which classes of galileid vacua can physically co-exist within the same EFT (if we are agnostic about galileid vacua in the first place). 
However, ideally we would like to do better and identify galileid vacua that are particularly `close' to the cosmological solution and hence the `correct' ones to use as theoretical priors for cosmology. To do so, recall that while a Galilean symmetry transformation $\phi \to \phi + c + c_\mu x^\mu$ can be used to shift $\phi$ and $\nabla_\mu \phi$, the second derivative $\nabla_\mu \nabla^\nu \phi$ is Galileon invariant. Comparing this for the cosmological background~\eqref{xiDef} with that of the vacua in Section~\ref{sec:2}, we see that, 
\begin{align}
&\text{Section~\ref{sec:2}:}  
&\nabla_\mu \nabla^\nu \bar{\phi} &= + \beta \Lambda_3^3 \left(  \delta_\mu^\nu - \left( 1  - \frac{\alpha}{\beta} \right) \delta_\mu^0 \delta^\nu_0  \right)    &( g_{\mu\nu} &= \eta_{\mu\nu} ) \; ,  \nonumber \\
&\text{Section~\ref{sec:3}:}
&\nabla_\mu \nabla^\nu \bar{\phi} &= - \xi \Lambda_3^3 \left( \delta_\mu^\nu - \left( 1 + \frac{\dot H}{H^2}  \right) \delta_\mu^0 \delta^\nu_0  \right)    &( g_{\mu\nu} &= g^{\rm FLRW}_{\mu\nu} )  \; .
\label{VacuaMapping}
\end{align}
%
\blue{We emphasise that both the background considered for the metric (a flat Minkowski background for section~\ref{sec:2} and a cosmological FLRW metric for section~\ref{sec:3}) and the scalar field background (the galileid vacua discussed in detail for section~\ref{sec:2} and a cosmological Galileon on the tracker discussed in detail above) are different for those two cases. However, \eqref{VacuaMapping} suggests that for sufficiently high energy $\varphi$ fluctuations (for whom $g^{\rm FLRW}_{\mu\nu}$ can be approximated as $\eta_{\mu\nu}$), an instantaneous mapping between the two solutions emerges at a given instant $t_i$ during the cosmological evolution, namely}
\begin{align} \label{betaXiMapping}
\beta &\sim -\xi, &\alpha/\beta &\sim - \dot H_i / H_i^2.
\end{align}
\blue{The scattering of sufficiently high energy $\varphi$ fluctuations can therefore instantaneously be described by the effective action of Section~\ref{sec:vacua} with $\{\alpha,\beta\}$ satisfying \eqref{betaXiMapping},
%
where we recall that $ \xi \equiv \dot\phi H/\Lambda_3^3$ is constant on the cosmologically relevant tracking solution, with its constancy (i.e. reaching the tracker before dark energy plays a significant role in the Universe's evolution) mandated by CMB constraints \cite{Barreira:2013jma}.
}

Equipped with the mapping \eqref{betaXiMapping}, we can therefore identify preferred positivity bounds for any instance in a given Galileon cosmology. We now investigate how these bounds can be used as theoretical priors in deriving cosmological parameter constraints. As before, we fix $c_2 = -1$ and solve for $c_4$ and $c_5$ using the Friedmann and current conservation equations. In addition we fix $\beta = -\xi$ from  \eqref{betaXiMapping} and use the tadpole condition for $(\alpha,\beta)$ vacua to find a unique corresponding $\alpha$. Overall this procedure associates a unique galileid $(\alpha,\beta)$ vacuum solution to a given $(\xi,c_3)$. Using \eqref{betaXiMapping} we can identify at which point in the evolution, i.e. for which time $t_i$ and $\dot H_i/H_i^2 = -\alpha/\beta$ this mapping establishes a direct link between the cosmological solution and the given $(\alpha,\beta)$ vacuum.
Note that, once we have fully solved for all parameters in terms of $(\xi,c_3)$, a specific choice of $\alpha/\beta$ traces out a line in the $(\xi,c_3)$ plane. For example, when choosing $\alpha/\beta = 3/2$ (i.e. a period of matter domination upon using \eqref{betaXiMapping}), then we find $c_3 \sim (3.8-1.2\xi^2)/\xi^3$. 

\begin{figure}[t!]
\begin{center}
\includegraphics[width=.328\linewidth]{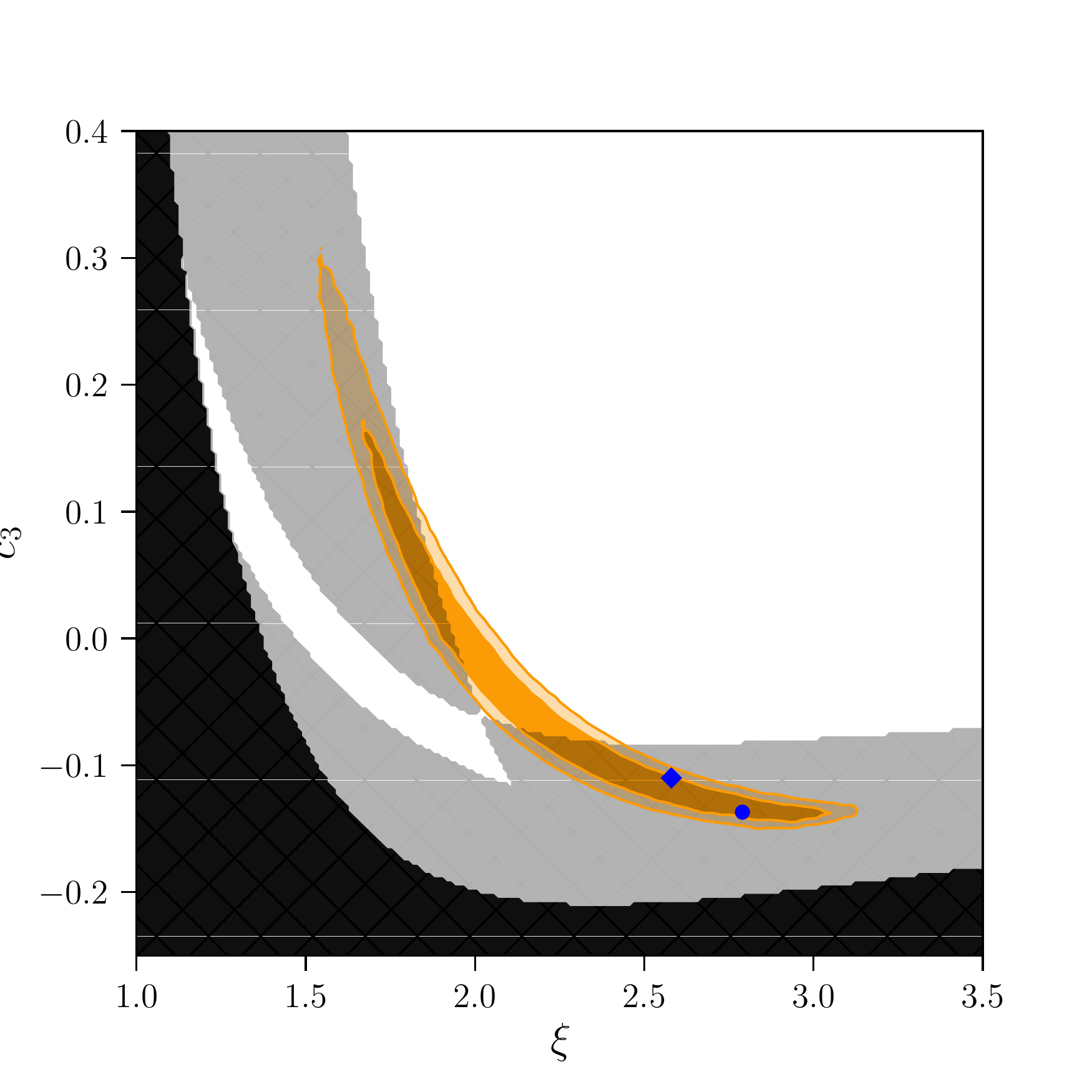}
\includegraphics[width=.328\linewidth]{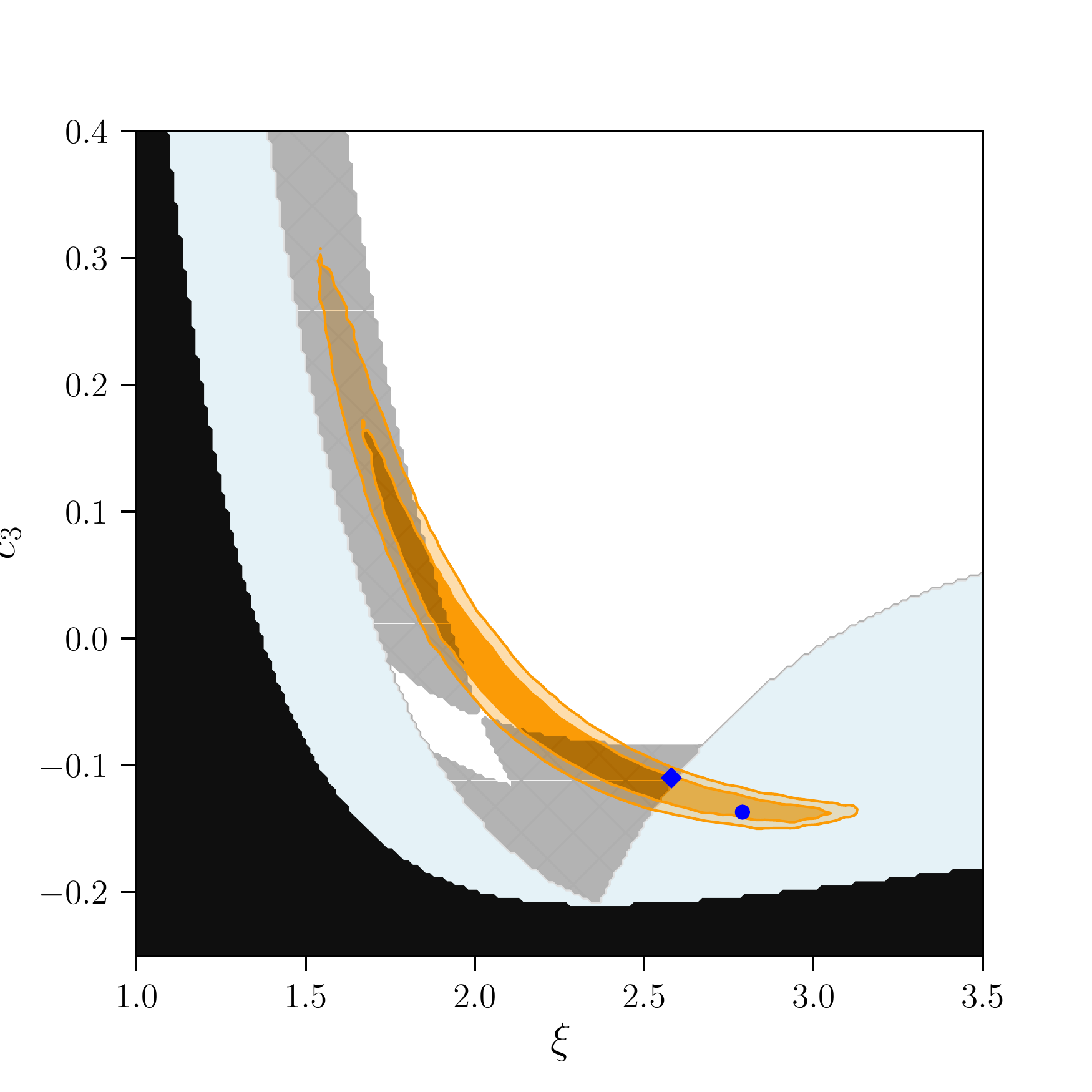}
\includegraphics[width=.328\linewidth]{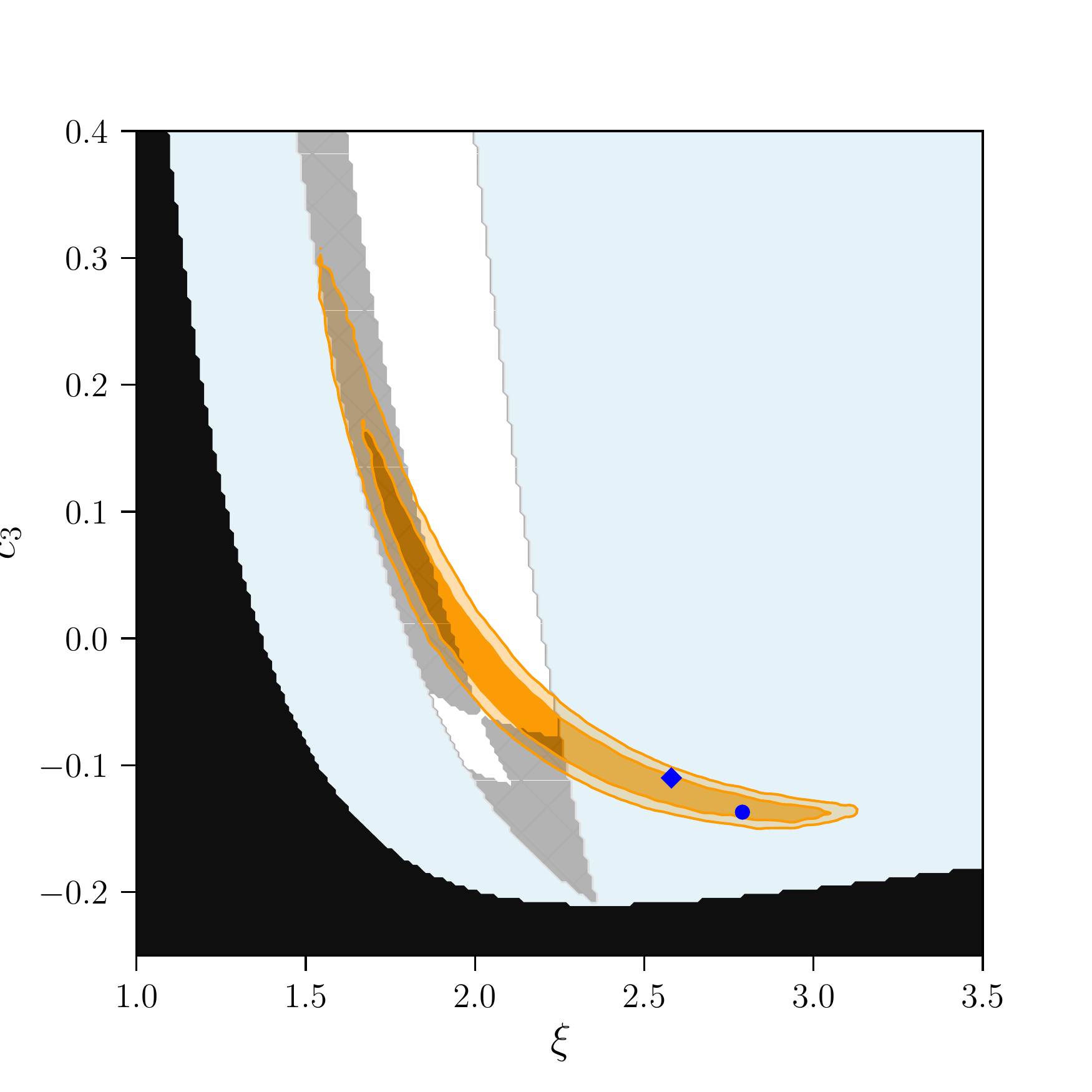}
\end{center}
\caption{
The equivalent of figure \ref{fig-regionPlot-boostBreak}, but here we fix $\beta = -\xi$ 
so that (together with the tadpole condition) there is a unique $(\alpha, \beta)$ vacuum assigned to each point in the $(\xi,c_3)$ plane. 
This assignment is such that, if $\dot H / H^2 \sim -\alpha/\beta$ at some stage during the cosmic evolution, then the cosmological $\bar \phi$ at that instant can be mapped to the corresponding 
$(\alpha, \beta)$ vacuum via \eqref{VacuaMapping} and \eqref{betaXiMapping}. 
This assigns a unique positivity bound to each point in the $( \xi, c_3 )$ plane, representing the bound from section~\ref{sec:2} which is most relevant for this particular cosmology. 
{\bf Left panel}: 
Points lying in hashed grey areas are ruled out by their respective positivity bounds, while black regions fail ghost/gradient stability tests. 
{\bf Middle panel}: Same as left, but with a blue shaded region, where $\dot H / H^2 = -\alpha/\beta$ is not achieved at any stage in the cosmic evolution from radiation domination to a future de Sitter limit, i.e. where $2 \geq -\dot H/H^2 = \alpha/\beta \geq 0$. So in the blue region the mapping \eqref{betaXiMapping} does not relate the $(\alpha, \beta)$ vacua assigned to individual points to instances during the actual cosmological evolution. Note that, under this mapping, specific values of $\dot H/H^2$ trace out lines in the panels shown.
The mapping \eqref{betaXiMapping} therefore cannot be invoked as robustly in the blue regions, but we can see that even only applying positivity priors outside those regions still excludes a large portion of the observationally relevant cosmologies (just about including one of the best-fit cosmologies).
{\bf Right panel}: Same as middle, but where blue shaded regions now denote parameter values for which $\dot H / H^2 = -\alpha/\beta$ is not achieved at any stage in the cosmic evolution from matter domination to today, i.e. where $3/2 \geq -\dot H/H^2 = \alpha/\beta \gtrsim 1/2$. 
This panel therefore displays a maximally conservative application of positivity bounds, with minimal extrapolation into the past and future. Even in this case positivity priors exclude a substantial fraction of the observationally relevant parameter space. Note that the vast majority of positive $c_3$ cosmologies is ruled out in all three cases shown.
\label{fig-regionPlotMapping}}
\end{figure}

The resulting constraints are shown in figure \ref{fig-regionPlotMapping}. The left panel shows the most direct application of the above procedure. Here we simply use~\eqref{betaXiMapping} to associate a specific $(\alpha,\beta)$ vacuum to any given point in the $(\xi,c_3)$ plane and compute and apply positivity bounds around that galileid vacuum regardless of the value of $\alpha/\beta$. One can then use those bounds as theoretical priors for cosmological parameter estimation and we see that there is significant tension between the region of parameter space preferred by the data themselves and positivity bounds here. About $2/3$ of the $2\sigma$ parameter space for $(\xi,c_3)$ identified by CMB+BAO constraints is inconsistent with the corresponding positivity requirements. Note that this includes the highlighted best-fit cosmologies, which prefer larger values for $\xi$. 
The middle and right panel then show more conservative applications of the identified positivity bounds. Here bounds from a given $(\alpha,\beta)$ vacuum identified via the above procedure are only applied, if the mapping \eqref{betaXiMapping} links them to a physical instance throughout cosmological evolution---regions where this is not the case are shaded blue and we do not place positivity priors in those regions. Specifically, the middle panel shows the case where we restrict to vacua with $2 \geq -\dot H/H^2 = \alpha/\beta \geq 0$ (i.e. from radiation domination to a future de Sitter limit). The right panel instead shows an even more conservative application where we restrict to vacua with $3/2 \geq -\dot H/H^2 = \alpha/\beta \gtrsim 1/2$ (i.e. from matter domination to today). Notably, whether we include radiation dominated phases or not only has a marginal impact on which regions in the $(\xi,c_3)$ plane we can constrain (slightly shrinking the region where we apply constraints from the left), but whether we extrapolate up to a future de Sitter limit or not has more significant implications (shrinking the region where we apply constraints from the bottom right. This future extrapolation is particularly important with an eye on constraining some of the observationally most-favoured large $\xi$ cosmologies. Note that (looking beyond the specific examples considered here), regardless of the specific family of vacua and/or form of the mapping used, establishing the range of cosmological evolution to which positivity bounds are applied is crucial in determining the resulting constraining power along the lines explored here. In particular, we expect that extrapolating up to a future de Sitter limit will always significantly increase constraining power in a dark energy context.

In all three cases shown in figure \ref{fig-regionPlotMapping} positivity bounds rule out most cosmologies with positive $c_3$. In the `medium' case we also (just) rule out the best-fit CMB+BAO+ISW cosmology and some of the region most preferred by data (namely the large $\xi$ cosmologies within the $2\sigma$ contours). The least conservative case also rules out the best-fit CMB+BAO cosmology and completely eliminates the large $\xi$ cosmologies ($\xi \gtrsim 2.5$) preferred by observational data alone.   
\\

\paragraph{Additional constraints from observations:} 
Throughout this paper we have only considered a small selection of observational bounds to illustrate the interplay between such constraints and theoretical positivity priors. In future work this can be improved upon by implementing/adding additional already existing theoretical bounds/observational data to the analysis. Here we point out a few potential examples for this route: 
\begin{itemize}
\item Incorporating galaxy-ISW cross-correlations into our analysis (cf. \cite{Kimura:2011td,Renk:2016olm,Renk:2017rzu,Lombriser:2016yzn}) as well as existing large scale structure measurements (e.g. from redshift space distortions or weak lensing observables) promises to add additional constraining power that would add to our analysis on linear cosmological scales. 
\item While we have restricted ourselves to considering the covariant Galileon as a dark energy EFT for cosmological scales here, 
complementary constraints can be obtained from significantly smaller scales probed by gravitational waves, e.g. by considering the propagation of dark energy perturbations on backgrounds sourced by binary SMBH mergers. This probes scales $\sim 10$ orders of magnitude smaller than cosmological scales and yields additional constraints on the size and presence of gravitational-wave induced dark energy instabilities \cite{Creminelli:2019kjy} (also see the closely related \cite{Creminelli:2019nok,Creminelli:2018xsv}). This in turn can significantly tighten cosmological parameter constraints on dark energy \cite{Noller:2020afd}, so investigating the interplay of these constraints will be an interesting task for the future. Here we have focused on observational constraints from cosmological scales only, in effect conservatively applying our EFT description of dark energy on cosmological scales only.
\item Finally, we have not incorporated any bounds on the speed of gravitational waves here. 
While especially the near simultaneous detections of GW170817 and GRB 170817A \cite{PhysRevLett.119.161101,2041-8205-848-2-L14,2041-8205-848-2-L15,2041-8205-848-2-L13,2041-8205-848-2-L12} have also been used to significantly reduce the functional freedom both for the covariant Galileon as well as for wider classes of dark energy/modified gravity theories (see \cite{Baker:2017hug,Creminelli:2017sry,Sakstein:2017xjx,Ezquiaga:2017ekz} and references therein), the frequencies of the merger are close to $\Lambda_3$, so additional assumptions about the UV physics are necessary to apply these bounds \cite{deRham:2018red}. Our goal here is to remain as agnostic as possible about the UV physics, so we will not fix the speed of cosmological gravitational waves here. Note that this is also in the spirit of only applying our EFT description of dark energy on cosmological scales, as mentioned above. 
\end{itemize}

\section{Discussion}
\label{sec:discussion}

\paragraph{Summary:}
We have explored how positivity constraints can vary when computing them around different vacuum solutions by considering the simple covariant Galileon model~\eqref{CovGal_action} as an instructive and illustrative example. 
These different positivity bounds can provide a valuable sign-posting of the IR landscape, identifying regions of parameter space in which one or more of the EFT vacua cannot possibly survive UV completion by any unitary, causal, local new physics. 
Particularly interesting are cases in which the usual (trivial) vacuum $\phi = 0$ is either not stable or violates the positivity bounds---we have shown that such theories may nonetheless be UV completed providing they possess a stable non-trivial field configuration which satisfies the corresponding bound. 
This is especially relevant in a dark energy context, where---both for the covariant Galileon we have focused on here as well as for large classes of other observationally relevant scalar-tensor theories---cosmological observations require that the kinetic term has the ``wrong sign'' for the trivial solution  \cite{Barreira:2013jma,Deffayet:2010qz,Traykova:2021hbr}, and so any comparison between positivity and cosmological data must employ bounds from other (non-trivial) solutions.
We showed that present observational constraints are incompatible with the existence of large families of such non-trivial solutions in the UV completion of the theory and we illustrated how this can be used to probe the underlying vacuum structure with present-day constraints from data and/or to tighten data constraints by imposing priors from the existence of well-motivated vacua. 
Finally, from the various background solutions for $\phi$ that we consider in this work, we identify one which is ``closest'' to the cosmological background evolution at a given time (in the sense that the scalar field profiles match instantaneously). Positivity bounds computed around this vacuum are therefore especially well-motivated as theoretical priors for cosmology and we have shown how the corresponding bounds impact cosmological parameter constraints, e.g. reducing the viable region of parameter space by up to $\sim 70 \%$
(with the precise percentage depending on the range of vacua from which the bounds are inferred)
and, for instance, ruling out the vast majority of otherwise observationally viable cosmologies with $c_3 > 0$.

~

These results---in particular the improved understanding of how different vacua lead to different positivity bounds---open up several new directions for future exploration.

\paragraph{Beyond the Covariant Galileon:}
To illustrate our main point, we have focussed on a particularly simple theory: the covariant Galileon. 
This had the advantage that there are many Lorentz-invariant and boost-breaking background solutions about which to derive positivity bounds and a simple tracker solution for the cosmological background evolution. 
In other more realistic models, it will generically be the case that expanding around different stable backgrounds will produce different positivity bounds, and therefore a more refined application of positivity bounds in the future will account for which subset of these vacua are assumed to persist in the UV theory.

\paragraph{EFT of Dark Energy:}
The time-dependence of the cosmological background is generally more complicated than simply $\phi \sim t^2$. 
One way to derive positivity bounds directly on this background would be to treat the scalar fluctuations as in the EFT of Dark Energy~\cite{Gubitosi:2012hu}. 
This shares many similarities with the EFT of inflation~\cite{Cheung:2007st}, and in particular in a suitable subhorizon/decoupling limit the corresponding scattering amplitudes would break only boosts (see for instance the discussion in~\cite{Grall:2021xxm}). This is potentially an avenue towards applying this positivity technology directly on the cosmological background (albeit on subhorizon scales). 
%

\paragraph{Improved Positivity Bounds:}
Recently there has been much progress developing positivity bounds on the scattering of scalar fields about Lorentz-invariant backgrounds. In particular, exploiting full crossing symmetry can lead to \emph{upper} bounds on coefficients like $c_{sst}$ \cite{Tolley:2020gtv, Caron-Huot:2020cmc, Sinha:2020win, Raman:2021pkf, Haldar:2021rri}, which complement the lower bounds considered here. 
Our observation that expanding around different vacua will correspond to different bounds on the EFT parameters remains true also for these additional positivity bounds.
It is worth remarking that these further bounds from full crossing symmetry have not yet been extended to boost-breaking backgrounds, and so for the $\alpha \neq \beta$ backgrounds considered above the only bounds currently available are the lower bounds shown here. 
It would also be interesting to further investigate the gravitational corrections to the positivity bounds studied here, for instance along the lines of \cite{Alberte:2020bdz, Alberte:2020jsk, Tokuda:2020mlf, Herrero-Valea:2020wxz, Caron-Huot:2021rmr, Alberte:2021dnj}.

\paragraph{Future data constraints:}
As discussed, the predictions of the covariant Galileon (as cosmological dark energy) are in some tension with the latest BAO data \cite{Renk:2017rzu}, e.g. with results from the  main galaxy sample (MGS) of SDSS DR7 \cite{Ross:2014qpa}. So future more precise BAO data have the potential to conclusively rule out this specific class of theories as well as to place significant constraints on dark energy effective field theories at large. Secondly, for the Galileon cosmologies we have focused on here, there is a strong $\sim 5\sigma$ preference for a non-zero sum of neutrino masses \cite{Renk:2017rzu}, more specifically a preference for a rather large sum of neutrino masses $\sum m_\nu \sim 0.5 \rm eV$. 
While the total neutrino mass is known to be at least $\sim 60$ meV from oscillation experiments \cite{Feldman:2012jdx,Mohapatra:2006gs,Gonzalez-Garcia:2002bkq,Maltoni:2004ei}, 
near-future experiments are forecast to {\it measure} this sum of masses with an error of $\sim 30$ meV \cite{Mishra-Sharma:2018ykh}, drastically improving constraints and placing significantly more precise bounds on (and potentially decisively ruling out) models such as the covariant Galileon and other EFTs of dark energy.
More generally, and going beyond the context of the specific illustrative models considered here, increasingly accurate measurements of clustering and lensing by large scale structure are expected to tighten current observational bounds (e.g. on the $c_i$  as considered here) by approx. an order of magnitude in the near future (LSST, CMB-S4, SKA) \cite{Alonso:2016suf}. Combined with novel constraints from gravitational waves, this should allow us to obtain far more precise hints from data as to the underlying nature of dark energy.

~

\comment{
\begin{figure}[t!]
\begin{center}
\includegraphics[width=.47\linewidth]{regplotMappingOmegaDiff.pdf}
\includegraphics[width=.52\linewidth]{posCovGal_scatterPlot.pdf}
\end{center}
\caption{
{\bf Left panel}: Differences in the `positivity exclusion regions' due to varying $\Omega_{\rm DE}$ within the $2\sigma$ constraints from Planck 2018, i.e. $\Omega_{\rm DE} = 0.6847 \pm 0.0146$. The two sets of contours shown here are for the two extreme values, i.e. $\Omega_{\rm DE} = 0.6993$ and $\Omega_{\rm DE} = 0.6701$. As can be seen from the figure, no significant difference on the final contours is induced by this change in $\Omega_{\rm DE}$. {\bf Right panel}: Scatter plot for the pure CMB+BAO data constraints for the quintic covariant Galileon, as also shown in figures 2 and 3. We can clearly see two `cuts'. First, around $c_3 = 0$ due to the solver degeneracy discussed above. Secondly at the lower right boundary of the contours - this sharp cut is triggered by the onset of gradient instabilities. 
\label{fig-OmegaPlot}}
\end{figure}
}

\subsubsection*{Acknowledgements}
We thank Claudia de Rham and Andrew Tolley for helpful discussions. 
Numerical computations were partially carried out on the Sciama High Performance Compute (HPC) cluster which is supported by the ICG, SEPNet and the University of Portsmouth.
SM is supported by an UKRI Stephen Hawking Fellowship (EP/T017481/1), and also partially by STFC consolidated grant ST/T000694/1.
JN is supported by an STFC Ernest Rutherford Fellowship (ST/S004572/1). 
In deriving the results of this paper, we have used: CLASS \cite{Blas:2011rf}, corner \cite{corner}, hi\_class \cite{Zumalacarregui:2016pph,Bellini:2019syt}, MontePyton \cite{Audren:2012wb,Brinckmann:2018cvx} and xAct \cite{xAct}.

\appendix
\section{Galileid interaction coefficients}
\label{app:galileid}

The expressions \eqref{Jcs} and \eqref{eqn:coefs2} given in the main text for the effective interaction coefficients of $\varphi$ around a general ($\alpha \neq \beta$) Galileid background are written in terms of the intermediate $\bar{c}_n$ coefficients \eqref{eqn:coefs1} of the Lorentz-invariant ($\alpha = \beta$) background. 
For ease of reference, note that in terms of the original $c_n$ coefficients appearing in the covariant Galileon action~\eqref{CovGal_action}, 
\begin{align}
 J &= c_1  - 2 ( \alpha + 3 \beta) c_2 - 24  \beta ( \alpha + \beta) c_3 - 
 24 \beta^2 (3 \alpha + \beta) c_4 + 16 \alpha \beta^3 c_5   \nonumber \\ 
  Z^2/c_s^3 &= c_2 + 12 \beta c_3 + 36 \beta^2 c_4 - 8 \beta^3 c_5   \\ 
 1- c_s^2 &=  \left( \beta -  \alpha  \right)  \frac{ 4 c_3 + 24 \beta c_4 - 8 \beta^2 c_5 }{ Z^2 / c_s^3 }   \, ,  \nonumber 
\end{align}
and
\begin{align}
\sqrt{-\tilde{g}} \, \tilde{c}_3 &= \frac{1}{c_s^4} \left(c_3^{\text{}} + 3\left(\alpha + \beta\right) c_4^{\text{}} - 2 \alpha \beta c_5^{\text{}} \right) \nonumber   \\
\sqrt{-\tilde{g}} \, \tilde{d}_3 &=  \frac{2}{c_s^4} \left( c_3 \left(1 -  c_s^2\right) +  3 c_4 \left(\alpha + \beta - 2\beta c_s^2\right) - 2c_5 \left( \alpha \beta - \beta^2 c_s^2\right)  \right)  \nonumber \\
\sqrt{-\tilde{g}} \, \tilde{c}_4 &= \frac{1}{c_s^6}\left( c_4 - \frac{2}{3} \alpha c_5 \right)  \nonumber \\
\sqrt{-\tilde{g}} \, \tilde{d}_4 &= \frac{1}{c_s^6} \left(3c_4 \left(1 -c_s^2\right) -2 c_5 \left(\alpha - \beta c_s^2\right)       \right)  \; . 
\end{align}

\section{Positivity bound details}
\label{app:positivity}

In this Appendix we collect various details that underpin the positivity bounds studied in the main text. 
First, in section~\ref{app:derivation} we briefly review the arguments that connect unitarity, causality and locality in the UV to bounds on the EFT coefficients. Then in section~\ref{app:unitarity} we compare these positivity bounds with the bounds which would have been inferred from perturbative unitarity alone. In particular, we show that for the covariant Galileon considered here, loop corrections to the EFT can remain small in the limit $\gamma \gg 1$ at small $s$ which was used in the main text to derive positivity bound \eqref{eqn:dt_pos2}. The role of the $\gamma$ parameter and its physical interpretation are described in more detail in section~\ref{app:gamma}. Finally, in section~\ref{app:noGalSymm} we explicitly write these positivity bounds for a general shift-symmetric amplitude (with no Galileon symmetry), to facilitate future applications to other scalar field theories.

\paragraph{$\bm{2 \to 2}$ scattering amplitude:}
The central object in deriving these UV/IR connections is the $2\to2$ elastic scattering amplitude, $\mathcal{A}$.
Since the general background \eqref{eqn:galileid} breaks 3 boost symmetries, the amplitude can depend on 5 independent variables (3 more than the usual Mandelstam $s$ and $t$), which we take to be $\mathcal{A} ( s , t ,  \omega_1,  \omega_2 , \omega_3 ) $ (since time translations are unbroken, the energy of particle 4 is fixed as $\omega_4 = -\omega_1 - \omega_2 - \omega_3$). 
Here $\omega_A$ is the time-like component (i.e. the frequency) of the 4-momentum $p^\mu_A$ for particle $A$.
Furthermore, since the propagator of $\varphi$ fluctuations is given by $( \tilde{g}^{\mu\nu} p_\mu p_\nu)^{-1}$ in momentum space, it is most natural to define the Mandelstam variables as,
\begin{align}
 s &:= - (p_1 + p_2)_\mu \tilde{g}^{\mu\nu} (p_1 + p_2 )_\nu
 \label{stu}
   \\ 
 t &:=  - (p_1 + p_3)_\mu \tilde{g}^{\mu\nu} (p_1 + p_3 )_\nu  \\
 u &:=  - (p_1 + p_4)_\mu \tilde{g}^{\mu\nu} (p_1 + p_4 )_\nu 
\end{align}
where our convention is that $\sum_a p_a = 0$ (and recall from section~\ref{sec:2} that $\tilde{g}^{\mu\nu} = \text{diag} \left( -1 , c_s^2 ,c_s^2 , c_s^2 \right)$). These obey the usual relation, $s+ t+ u =0$, since on-shell fluctuations satisfy the free equation of motion\footnote{
Or rather $p_{ \mu} \tilde{g}^{\mu\nu} p_{ \nu}=\mathcal{O}(m^2)$, if the shift symmetry is softly broken by a small mass.
} $p_{ \mu} \tilde{g}^{\mu\nu} p_{ \nu}=0$.
In terms of these variables, the $2 \to 2$ scattering amplitude from the cubic and quartic interactions \eqref{eqn:Scubic} and \eqref{eqn:Squartic} is,
\begin{align}
 \mathcal{A} (s, t, \omega_1, \omega_2, \omega_3 ) =& 
\left(  \tilde{c}_4 -  \frac{\tilde{c}_3^2}{Z^2} \right)  \frac{ s^3 + t^3 + u^3 }{\Lambda_3^6}   \nonumber \\
&+\left( \frac{ 2 \tilde{c}_3 \tilde{d}_3 }{Z^2} -  \tilde{d}_4 \right) \frac{ s^2 ( \omega_1 \omega_2 + \omega_3 \omega_4) + 
  t^2 ( \omega_1 \omega_3 + \omega_2 \omega_4 ) +
  u^2 ( \omega_1 \omega_4 + \omega_2 \omega_3 )    }{ \Lambda_3^6}  
  \label{eqn:A}
\end{align} 
where we note that the $\tilde{d}_3^2$ term is proportional to $s+t+u$ and hence vanishes.

\subsection{Overview of positivity derivation}
\label{app:derivation}

In this section we briefly review the arguments of \cite{Adams:2006sv}, \cite{deRham:2017avq} and \cite{Grall:2021xxm} which connect the properties of unitarity, causality (analyticity) and locality (boundedness) of the underlying UV theory to the bounds \eqref{eqn:posLO_css}, \eqref{eqn:posNLO_csst} and (\ref{eqn:posLO_css2}, \ref{eqn:posNLO_csst2}) on the low-energy EFT coefficients.

\paragraph{Lorentz-invariant bounds:}
For Lorentz-invariant interactions, the scattering amplitude is a function of the two Mandelstam variables only, $\mathcal{A} (s,t)$. 
Assuming a unitary, causal, local (Lorentz invariant) UV completion leads to various positivity bounds on $\mathcal{A} (s,t)$. In short, unitarity (namely a partial wave expansion of the optical theorem) can be used to show that\footnote{
Note that $\partial_t^n \text{Im} \mathcal{A} (s,t) |_{t=0} = 0$ requires that $\mathcal{A} = 0$ at all energies, i.e. that the theory can have no interactions and is trivially free.  
},
\begin{align}
 \partial_t^n \, \text{Im} \mathcal{A} (s, t) |_{t=0} > 0 
 \label{eqn:unitarity}
\end{align}
for any physical value of $s$ and any number of $t$ derivatives, while causality and locality can be used to prove a simple dispersion relation which relates the amplitude at different scales,
\begin{align}
 \partial_s^2 \mathcal{A}  ( s, t ) =  \int_{0}^{\infty} \frac{d s'}{\pi}  \; \text{Im} \, \mathcal{A} ( s' ,t) \left[  \frac{1}{( s' - s )^3} +   \frac{1}{( s'  + s + t  )^3}  \right]     \; ,
 \label{eqn:dispersion}
\end{align}
where we have neglected the particle masses\footnote{
In practice a gapped theory will have $\text{Im} \mathcal{A} = 0$ for $0 < s' < 4m^2$ below the two-particle threshold, and also $| \mathcal{A} | < s^2$ at large $s$ due to the Froissart bound. Although we will neglect mass effects, we continue to assume this Froissart boundedness in the UV.
}, and made use of the crucial crossing property that $\mathcal{A} (s, t) = \mathcal{A} (u,t)$.  
\eqref{eqn:dispersion} can be used to connect unitarity \eqref{eqn:unitarity} in the UV (namely positivity of $\text{Im} \, \mathcal{A} (s', t)$ at large $s'$) to properties of the low-energy EFT (namely the sign of $\partial_s^2 \mathcal{A} (s,t)$ at low $s$). 
Explicitly, supposing that the low-energy EFT can be used to reliably compute $\mathcal{A}$ up to an energy $s_b$ (beyond which it breaks down and is replaced by a UV completion), then we can use \eqref{eqn:dispersion} to relate $\mathcal{A}_{\rm EFT}$ and $\mathcal{A}_{\rm UV}$,
\begin{align}
 \partial_s^2 \mathcal{B}_{\rm EFT} ( s, t ) &:= \partial_s^2 \mathcal{A}_{\rm EFT} -  \int_{0}^{s_b} \frac{d s'}{\pi}  \; \text{Im} \, \mathcal{A}_{\rm EFT} ( s' ,t) \left[  \frac{1}{( s' - s )^3} +   \frac{1}{( s' + s + t )^3}  \right]   \nonumber \\
 &\;=  \int_{s_b}^{\infty} \frac{d s'}{\pi}  \; \text{Im} \, \mathcal{A}_{\rm UV} ( s' ,t) \left[  \frac{1}{( s' - s )^3} +   \frac{1}{( s' + s + t )^3}  \right]     \; ,
\end{align}
where we have introduced $\mathcal{B}_{\rm EFT}$ as shorthand for the low-energy part of \eqref{eqn:dispersion}.
A unitary, causal and local UV completion at high energies therefore requires various bounds on $\mathcal{A}_{\rm EFT}$---in particular the forward limit must be positive \cite{Adams:2006sv},
\begin{align}
 \partial_s^2 \mathcal{B}_{\rm EFT}  |_{\substack{s=0 \\ t=0 }} &=  \frac{2}{\pi} \int_{s_b}^{\infty} \frac{ds}{s^3}  \; \text{Im} \, \mathcal{A}_{\rm UV} |_{t=0}   > 0     \; , \label{eqn:posLO}   
\end{align}
and the $t$ derivatives must be bounded in terms of lower-order derivatives \cite{deRham:2017avq}, 
\begin{align}
 \left( \partial_t + \tfrac{3}{2 s_b}   \right) \partial_s^2 \mathcal{B}_{\rm EFT}  |_{\substack{s=0 \\ t=0 }} &=  \frac{2}{\pi} \int_{s_b}^{\infty} \frac{ds}{s^3}  \left( \partial_t + \tfrac{3}{2} \tfrac{s - s_b}{s \, s_b}   \right) \text{Im} \, \mathcal{A}_{\rm UV} |_{t=0}   > 0  \; , \label{eqn:posNLO} 
\end{align}
and so on. When applied in perturbation theory, where the EFT amplitude takes the form~\eqref{eqn:Aexp}, \eqref{eqn:posLO} implies \eqref{eqn:posLO_css} and \eqref{eqn:posNLO} implies \eqref{eqn:posNLO_csst}.

\paragraph{Boost-breaking bounds:}
For a general amplitude $\mathcal{A} (s,t, \omega_1 ,\omega_2, \omega_3)$ in which boosts are broken, unitarity (namely a suitable spherical wave expansion of the optical theorem) can again be used to establish that \cite{Grall:2020tqc},
\begin{align}
\partial_t^n \text{Im} \, \mathcal{A} (s, t, \omega_1 ,\omega_2, \omega_3 ) |_{ \substack{t=0 \\ \omega_1 = - \omega_3} } > 0 
\label{eqn:unitarity2}
\end{align}
for any physical value of $s$ and of the energies, where now the forward limit corresponds to both $t=0$ and $\omega_1 = - \omega_3$. 
Translating \eqref{eqn:unitarity2} into EFT positivity bounds was recently considered in \cite{Grall:2021xxm} (see also \cite{Baumann:2015nta} for earlier work with particular centre-of-mass kinematics). The central distinction with Lorentz-invariant positivity bounds is that some prescription must be provided for how to hold the three energy variables fixed when performing the partial derivatives and integration in any dispersion relation \eqref{eqn:dispersion}. 
In particular for $s$-channel scattering, since the spatial momenta $c_s | \mathbf{p}_1 + \mathbf{p}_2 | > \omega_1 - \omega_2$ on-shell, the Mandelstam $s$ defined in \eqref{stu} must obey,
\begin{align}
 s \leq ( \omega_1 + \omega_2 )^2 - (\omega_1 - \omega_2 )^2 \; .
 \label{sbound}
\end{align}
This means that holding $\omega_1$ and $\omega_2$ fixed is not an option, since \eqref{sbound} would always be violated at sufficiently large $s$, invalidating the unitarity bound on the UV amplitude (which only applies to physical on-shell momenta). Instead, it was argued in  \cite{Grall:2021xxm} that the correct prescription for the forward limit amplitude is,
\begin{align}
\tilde{\mathcal{A}}  ( s , \gamma , M )  := \mathcal{A} \left( s, \;  t=0,  \; \omega_1 = - \omega_3 = \gamma^2  M, \; \omega_2 = \tfrac{s}{4M}  \right)  \; , 
\label{eqn:Atilde}
\end{align}
where $\omega_1$ and $\omega_2$ have been replaced with the variables $\gamma^2$ and $M$, and the $s$-channel condition \eqref{sbound} becomes simply $M  > 0$ and $\gamma^2  \geq 1$.
Fixing $\omega_2$ in terms of $s$ in this way ensures that when $s$ is analytically continued to $-s$, $\omega_2$ is analytically continued to $-\omega_2 =  \omega_4$, and so one may retain a simple crossing relation, $\tilde{\mathcal{A}} ( s, \gamma , M ) =  \tilde{\mathcal{A}} ( -s , \gamma , M )$ (since particles 2 and 4 are indistinguishable). 
Following the same steps as in the Lorentz-invariant case, assuming the analogous properties (analyticity, crossing and Froissart boundedness\footnote{
In the Lorentz-invariant case, these properties are consequences of unitarity and causality in any local quantum field theory with a mass gap. When boosts are broken, there is not yet a rigorous derivation of these properties  from causality, however see \cite{Grall:2021xxm} for first steps in that direction.  
}) of $\tilde{\mathcal{A}} (s, \gamma , M)$ in the full UV theory leads to a dispersion relation of the form \eqref{eqn:dispersion} which connects the amplitude on different scales, allowing \eqref{eqn:unitarity2} to constrain the EFT amplitude.

The analogue of the leading positivity bound \eqref{eqn:posLO} when boosts are broken is,
\begin{align}
 \partial_s^2 \tilde{B}_{\rm EFT} |_{s=0} = \frac{2}{\pi}  \int_{s_b}^{\infty} \frac{ds}{s^3}  \; \text{Im} \, \tilde{\mathcal{A}}_{\rm UV}   > 0  \; ,
 \label{eqn:posLO2}
\end{align}
where the derivatives/integrals are taken with $\gamma, \, M$ held fixed, and the inequality holds for any $M > 0$ and $\gamma \geq 1$. 

To go beyond the forward limit, one might imagine that,
\begin{align}
 \tilde{\mathcal{A}} ( s , t, \gamma, M ) := \mathcal{A} \left( s, \;  t,  \; \omega_1 = -\omega_3 = \gamma^2  M, \; \omega_2 = \tfrac{s-u}{8M}  \right)
\end{align}
is the natural extension of \eqref{eqn:Atilde}, since this would retain a trivial crossing relation $\tilde{\mathcal{A}} (s,t,\gamma,M) = \tilde{\mathcal{A}} (u,t,\gamma,M)$. However, the $t$ derivatives, 
\begin{align}
 \partial_t \text{Im} \, \tilde{\mathcal{A}}   = \left( \partial_t + \tfrac{1}{8M} \partial_{\omega_2} \right) \text{Im} \, \mathcal{A}  \; . 
\end{align}
are \emph{not} positive, due to the $\partial_{\omega_2}$ term.
As described in the Appendix of \cite{Grall:2021xxm}, the key to going beyond the forward limit is to consider an \emph{integral} of the amplitude\footnote{
The integration limits have been chosen so that $\mathcal{I} ( u, t, \gamma, M , \delta) = \mathcal{I} ( s, t, \gamma, M , \delta)$ inherits the crossing relation of $\mathcal{A}$. 
},
\begin{align}
\mathcal{I} ( s, t, \gamma , M , \delta ) := \int_{ \frac{s-u}{8M} - \delta }^{\frac{s-u}{8M} + \delta } d \omega_2 \; \mathcal{A} ( s, t , \omega_1 , \omega_2 ,  \omega_3 ) |_{\omega_1 = - \omega_3 = \gamma^2 M}  \; ,
\end{align}
where we assume that the constant $\delta$ can be chosen sufficiently small that this integral converges for any $s$ (which is certainly the case in perturbation theory, since  $\mathcal{A} (s, t, \omega_1 , \omega_2, \omega_3 )$ is analytic in $\omega_2$ at fixed $s$), and provides a new complex function which shares the analyticity properties of $\tilde{\mathcal{A}}$. 
Unitarity \eqref{eqn:unitarity2} now guarantees that,
\begin{align}
\left( \partial_t + \tfrac{1}{8M} \partial_\delta   \right) \text{Im}\,  \mathcal{I} (s, t, \gamma , M , \delta ) |_{t=0} >  0  
\end{align}
for all $s > 0 , \;  M >0  , \; \gamma^2 \geq s / (s - 4 \delta M) \geq 1$ (to be compatible with the condition \eqref{sbound} for real momenta in the $s$-channel), and so a dispersion relation for $\mathcal{I}$ can be used to place a positivity bound on the EFT analogous to \eqref{eqn:posNLO},
\begin{align}
\left[ \partial_t + \tfrac{3}{2 s_b} + \tfrac{1}{8M} \partial_\delta   \right] \partial_s^2 \mathcal{I}_{\rm EFT} |_{\substack{s=0 \\ t=0 }}  
= \frac{2}{\pi} \int_{s_b}^{\infty} \frac{d s}{ s^3}  \;\left( \partial_t + \tfrac{1}{8M} \partial_\delta  + \tfrac{3}{2} \tfrac{ s - s_b}{ s \, s_b}  \right)  \text{Im} \, \mathcal{I}_{\rm UV}  |_{t=0}  
> 0 
\label{eqn:posNLO2}
\end{align}
which must be satisfied for all $\gamma^2 \geq 1 + \delta/\omega_b \geq 1$, where $\omega_b$ is the energy scale up to which the branch cut can be subtracted within the EFT. 
When applied in perturbation theory, where the EFT amplitude takes the form~\eqref{eqn:Aexp2}, \eqref{eqn:posLO2} implies \eqref{eqn:posLO_css2} and \eqref{eqn:posNLO2} implies \eqref{eqn:posNLO_csst2}.

\paragraph{EFT regime of validity:}
Our leading-order, tree-level computation of $\mathcal{A}_{\rm EFT} (s,t, \omega_1, \omega_2, \omega_3)$ is subject to corrections from (i) higher-derivative interactions in the EFT (which capture loops of all heavy fields integrated out of our description), and (ii) loops of the light fields. 
To estimate the size of (i), we appeal to a simple EFT power-counting in which all fields/derivatives are suppressed by the same cut-off scale $\Lambda$, 
\begin{align}
\mathcal{A}_{\rm EFT} 
= \mathcal{A}_{\rm LO}  + \mathcal{O} \left( \frac{s}{\Lambda^2} , \frac{\omega_1}{\Lambda} , \frac{\omega_2}{\Lambda}  \right) \; .
\end{align}
In order for our leading-order calculation $\mathcal{A}_{\rm LO}$ (which included only the interactions with fewest derivatives, namely \eqref{CovGal_action}) to be a reliable estimate, we require that,
\begin{align}
\frac{s^2}{\Lambda^2} \ll 1 \;\; , \;\; \frac{\omega_1}{\Lambda} \ll 1 \;\; , \;\; \frac{\omega_2}{\Lambda} \ll 1 \; .
\label{eqn:power_counting}
\end{align}
In particular, when applying the positivity bounds above, in which $\omega_1$ and $\omega_2$ are fixed in terms of $(s, \gamma, M)$, \eqref{eqn:power_counting} implies that higher-derivative corrections are only small providing that, 
\begin{align}
\frac{s}{\Lambda^2} \ll \frac{M}{\Lambda} \ll \frac{1}{\gamma^2} \leq 1 \; . 
\end{align}
Put another way, when $\gamma^2$ is large, this lowers the effective cut-off of our EFT in the complex $s$-plane to $s_{\rm max} \sim  \Lambda^2 / \gamma^2$. 
In the main text, when we discuss the large $\gamma^2 \gg 1$ limit, we have in mind that we evaluate $\mathcal{A}_{\rm EFT}$ at sufficiently small $s$ ($<s_{\rm max}$) that these higher-derivative corrections remain small. 

To estimate the size of (ii), the correction due to light loops, we can use perturbative unitarity. This is described in more detail in the following section, but in summary: since perturbative unitarity imposes a condition of the form $\text{Im} \, \mathcal{A} > | \mathcal{A} |^2$, we can use our tree-level calculation of $|\mathcal{A}|$ to place a lower bound on the size of the light loops (which contribute to $\text{Im} \, \mathcal{A}$). We find that the soft behaviour of the Galileon amplitude, $\lim_{s \to 0} \mathcal{A} = 0$, guarantees that there is never any obstruction to making $\gamma^2$ large, providing $s$ is made small enough\footnote{
In contrast, in the absence of any shift symmetry, one could have contributions to $\mathcal{A}$ from $s$-channel exchange that $\sim \omega_1^n/s = \gamma^{2n} M^n / s$, and these will violate perturbative unitarity at large $\gamma$ / small $s$.  
}.

\subsection{Comparison with perturbative unitarity bounds}
\label{app:unitarity}

In this section, we compare the positivity bounds studied in the main text with the bounds that can be placed on the EFT coefficients from perturbative unitarity alone. 
In short, unitary time evolution is encoded in the amplitude via the optical theorem,
\begin{align}
\frac{1}{i} \left( \mathcal{A}_{p_1 p_2 \to p_3 p_4} - \mathcal{A}_{p_3 p_4 \to p_1 p_2}^* \right) = \int_{q_1 q_2} \mathcal{A}_{p_1 p_2 \to q_1 q_2} \mathcal{A}_{p_3 p_4 \to q_1 q_2}^* + ... 
\label{eqn:optical}
\end{align}
where the $+...$ terms involve higher $n$-point amplitudes and are positive for forward scattering. 
The integral on the right-hand-side is over the two-particle phase space, subject to the momentum conservation condition that $p_1 + p_2 = q_1 + q_2$. 

\paragraph{Lorentz-invariant bounds:}
The partial wave expansion\footnote{
Note that here, as in the main text, we are treating all particle masses as negligible. 
}, 
\begin{align}
\mathcal{A}_{p_1 p_2 \to p_3 p_4} = 16 \pi g_s  \sum_{\ell} (2 \ell +1 ) P_{\ell} ( \cos \theta_s ) \; a_{\ell} (s )
\label{eqn:pw}
\end{align}
replaces $t$ with the centre-of-mass frame scattering angle, $\cos \theta_s = 1 + 2t/s$. 
The degeneracy factor $g_s$ accounts for whether the two ingoing particles are distinguishable ($g_s = 1$) or indistinguishable ($g_s = 2$) in the $s$-channel. 
Substituting \eqref{eqn:pw} into \eqref{eqn:optical} gives $2 \, \text{Im} \, a_{\ell} (s) = | a_{\ell} (s) |^2 + ...$, where the higher-order $+...$ terms are all positive. Unitarity can therefore be expressed in terms of the partial wave coefficients as, 
\begin{align}
1 \geq |a_{\ell} (s) |  \geq   \text{Im} \, a_{\ell} (s) \geq 0 \; . 
\label{eqn:pw_unitarity}
\end{align}

For the amplitude \eqref{eqn:Aexp}\footnote{
More precisely, from the $c_{ss}$ and $c_{sst}$ terms appearing in the full amplitude \eqref{eqn:ALO} and \eqref{eqn:ANLO}. 
}, the only non-zero partial wave amplitudes at tree-level are,
\begin{align}
a_0 (s) &= \frac{s^2}{192 \pi} \left( 5 c_{ss} - c_{sst} s \right)  \nonumber \\
a_2 (s) &= \frac{s^2}{960 \pi} \left( c_{ss} + c_{sst} s \right) \; . 
\label{eqn:a0a2}
\end{align}
Since $c_{ss} \sim 1/\Lambda^4$ and $c_{sst} \sim 1/\Lambda^6$ in terms of the EFT cut-off, the unitarity bound \eqref{eqn:pw_unitarity} is always satisfied in perturbation theory providing $s < \kappa \Lambda$, where $\kappa$ is an order one number (whose precise value can be inferred from \eqref{eqn:a0a2} if so desired). 

\paragraph{Boost-breaking bounds:}
When boosts are broken, the amplitude depends explicitly on the energies of the particles (as discussed in the previous section). 
One may nonetheless restrict one's attention to scattering processes for which $\mathbf{p}_1 + \mathbf{p}_2$ happens to vanish, i.e. those for which the partial wave expansion~\eqref{eqn:pw} may be used just as in the Lorentz invariant case. This corresponds to setting all energies $|\omega_n| = \sqrt{s}/2$. 
The resulting optical theorem can be used to place constraints on the EFT Wilson coefficients (see for instance \cite{Baumann:2011su, Baumann:2014cja}). 
However, to leverage the full constraining power of unitarity, we must allow for scattering at arbitrary kinematics, as described in \cite{Grall:2020tqc}.
This requires expanding the amplitude in terms of more general angular momentum states, and leads to the spherical wave expansion\footnote{
The 2-particle partial wave expansion in a general Lorentz frame can also be found in \cite{Wick:1962zz}.
},
\begin{align}
\mathcal{A}_{p_1 p_2 \to p_3 p_4} = 64 \pi^2 g_s \sum_{ \substack{\ell_1 \ell_3 \\ m_1 m_3} } Y_{\ell_1}^{m_1} ( \vartheta_1 , \varphi_1 ) \, Y_{\ell_3}^{m_3 *} (\vartheta_3 , \varphi_3 ) \, a^{m_1 m_3}_{\ell_1 \ell_3} ( s , \gamma_s ) \; ,
\label{eqn:sw}
\end{align}
where the angles $\vartheta_n$ and $\varphi_n$ describe the direction of $\mathbf{p}_n$ (relative to $\mathbf{p}_1 + \mathbf{p}_2$), and $\gamma_s$ is the Lorentz factor associated with the centre-of-mass motion, i.e. $\gamma_s = 1/\sqrt{  1 - v_{\rm CM}^2/c_s^2 }$ where $v_{\rm CM} = c_s^2 | \mathbf{p}_1 + \mathbf{p}_2 |/(\omega_1 + \omega_2)$. 
These are explicitly related to the variables used in the main text by\footnote{
Note that the amplitude does depends only on the difference $\varphi_1 - \varphi_3$ (since rotations about $\mathbf{p}_1 + \mathbf{p}_2$ remain unbroken by the centre-of-mass motion), and consequently $a_{\ell_1 \ell_3}^{m_1 m_3} ( \omega_s, s)$ is diagonal in its $m_1$ and $m_3$ indices. 
Further selection rules on these spherical wave coefficients can be found in the Appendix of \cite{Grall:2021xxm}. 
},
\begin{align}
|\omega_n| &= \frac{\sqrt{s}}{2}  \, \gamma_s \left( 1 -   \cos \vartheta_n \right) \; ,    \nonumber \\
t &=  \frac{s}{2} \left( 
- 1 + \cos \vartheta_1 \cos \vartheta_3 + \sin \vartheta_1 \sin \vartheta_3 \cos \left( \varphi_1 - \varphi_3 \right)  
\right)    \;  .  
\end{align}
Note that this $\gamma_s$ is related to the $\gamma$ appearing in the positivity bounds, but they are not identical---the precise relation between the two is given in the next section. 
Substituting \eqref{eqn:sw} into \eqref{eqn:optical}, unitarity can be expressed in terms of the spherical wave coefficients as, 
\begin{align}
\left|  \sum_{\substack{\ell_1 \ell_3 \\ m_1 m_3 }}  v^{m_1}_{\ell_1} a_{\ell_1 \ell_3}^{m_1 m_3} ( s, \gamma_s )  v^{m_3 *}_{\ell_3}  \right|  \leq  \sum_{\substack{\ell \\ m }}  v^m_{\ell}  v^{m *}_{\ell}  \; . 
\label{eqn:sw_unitarity}
\end{align}
for any complex vector $v^{m}_{\ell}$, which is the matrix analgoue of $| a_{\ell} | \leq 1$. 

For the Galileon amplitude \eqref{eqn:Aexp2}\footnote{
More precisely, from the Galileon-invariant terms appearing in the full amplitude \eqref{eqn:ANLO}. 
}, the non-zero spherical wave amplitudes are\footnote{
Note that the centre-of-mass frame corresponds to $\gamma_s = 1$. With this choice, the spherical wave coefficients are related to the traditional partial wave coefficients by,
$a_{\ell} (s) = \frac{1}{2 \ell +1} \sum_{m=-\ell}^{\ell} a^{mm}_{\ell \ell} (s , \gamma_s ) |_{\gamma_s = 1} $.
},
\begin{align}
a^{00}_{00} &= \frac{s^3}{192 \pi} \left[ - c_{sst} - \frac{1}{6} d_{st\omega\omega}  - \frac{\gamma_s^2}{12} d_{st\omega\omega}  \right]    \nonumber \\
a^{00}_{02} &= a^{00}_{20} =   \frac{s^3}{2304 \sqrt{5} \pi}  \left[ d_{st\omega\omega} \left( \gamma_s^2 - 1 \right)  \right]  \nonumber \\
a^{00}_{22}  &= \frac{s^3}{960 \pi} \left[  c_{sst} + \frac{1}{3} d_{st\omega\omega}  - \frac{\gamma_s^2 }{12} d_{st \omega\omega}  \right] \nonumber \\
a^{\pm 1, \pm 1}_{22} &=   \frac{ s^3 }{ 960 \pi } \left[    c_{sst}  + \frac{1}{4} d_{st \omega\omega}  \right]   \nonumber \\ 
a^{\pm 2, \pm 2}_{22} &=  \frac{s^3}{960 \pi} \left[   c_{sst}  +  \frac{\gamma_s^2}{4} d_{st \omega\omega}   \right]  \; . 
\label{eqn:sw_eg}
\end{align}
Since $c_{sst}$ and $d_{st\omega\omega}$ both $\sim 1/\Lambda^6$ in terms of the EFT cut-off, the unitarity bound \eqref{eqn:pw_unitarity} is always satisfied in perturbation theory providing $\gamma_s^2 s^3 < \kappa' \Lambda^6$, where $\kappa'$ is again an order one number (which can be inferred from \eqref{eqn:sw_unitarity} if so desired). 

The important point here is that, while perturbative unitarity bounds like \eqref{eqn:pw_unitarity} and \eqref{eqn:sw_unitarity} are useful for estimating the regime of validity of the EFT, they invariably constrain the absolute size of Wilson coefficients, but not their overall sign. The positivity bounds discussed in the main text however, since they are exploiting causality and locality in addition to unitarity, are able to say something non-trivial about the signs of each coefficient appearing in the amplitude. 

Finally, note that \eqref{eqn:sw_eg} confirms that, providing the interaction energy $s$ is sufficiently small, the centre-of-mass velocity may be large (i.e. $\gamma_s \gg 1$), without generating large loop corrections (i.e. loops can be small without violating the bound \eqref{eqn:sw_unitarity}).
Since when $s$ is small, $\gamma_s \sim \gamma^2 M /\sqrt{s}$, this also implies that taking $\gamma \gg 1$ at fixed $M$ and $s$ (as we do in the main text to derive \eqref{eqn:dt_pos2}) can remain safely within the EFT's regime of validity. 
We will now discuss the link between $\gamma^2$ and $\gamma_s^2$ in more detail.

\subsection{The role of $\gamma^2$}
\label{app:gamma}

In this short section, we describe the physical interpretation of the parameter $\gamma$ that appears in the positivity bounds~\eqref{eqn:posLO_css3} and \eqref{eqn:posNLO_csst2}. 
It is related to the Lorentz factor $\gamma_s$ of the centre-of-mass motion, however $\gamma_s$ is not a manifestly crossing symmetric quantity. In order for the amplitude to exhibit an $s\leftrightarrow u$ crossing symmetry, we should instead hold fixed a combination of $\gamma_s$ and $\gamma_u$ (its $u$-channel counterpart) which is manifestly $s \leftrightarrow u$ symmetric. 
One natural choice is, 
\begin{align}
 \gamma^2 =  \frac{ s \gamma_s^2 - u \gamma_u^2 }{s-u}  = \gamma_s^2 + \gamma_u^2 \;\; \text{when } t=0
\end{align}
where,
\begin{align}
\gamma_s^2 = 1 + \frac{c_s^2 | \mathbf{p}_1 + \mathbf{p}_2 |^2 }{s} \; , \;\;\;\; \gamma_u^2 = 1 + \frac{c_s^2 | \mathbf{p}_1 + \mathbf{p}_4 |^2 }{u}
\end{align}
represent the ``centre-of-mass motion'' in the $s$- and $u$-channels respectively.
This particular combination corresponds to the velocity of the so-called Breit frame, in which the analytic structure of the amplitude is simplest \cite{Grall:2021xxm}.
Note that for the physical scattering process $p_1 p_2 \to p_3 p_4$, $\gamma_u$ does not represent the speed of any physical object (in particular $\gamma_u^2 < 0$), but rather is the analytic continuation of $\gamma_s$ under $p_2 \leftrightarrow p_4$. 
In fact, if one focuses on real momenta in the $s$-channel, then in the forward limit one can write $\gamma^2 = \cos^2 \vartheta_1 + \gamma_s^2 \sin^2 \vartheta_1$, in terms of the angular variables of the previous section---in particular, $\gamma \leq \gamma_s$ and is only $=\gamma_s$ when $\vartheta_1$ approaches $\pi/2$.  

At this point, it is worth noting that there is a caveat to the second positivity bound~\eqref{eqn:dt_pos2}.
A violation of \eqref{eqn:dt_pos2} could be consistent with a unitary, causal, local UV completion if the EFT breaks down before a maximum value of $\gamma^2$, 
\begin{align}
\gamma_{\rm max}^2  =  \frac{ \tilde{c}_3^2  - Z^2 \tilde{c}_4 }{  \tfrac{2}{3} \tilde{c}_3 \tilde{d}_3 -\tfrac{1}{3} Z^2 \tilde{d}_4 } \; ,
\label{eqn:gmax}
\end{align}
even at $s \approx 0$. 
Physically, this corresponds to an EFT which cannot resolve large centre-of-mass motions with $\gamma_s  \geq \gamma_{\rm max}$, no matter how small the interaction energy $s$ is.  
As we argued above, there can be no such breakdown in $\gamma$ if integrating out the heavy physics leads to a simple EFT power-counting in which all fields/derivatives are suppressed by the same scale. But in principle a more exotic UV theory, which produces a low-energy EFT in which temporal/spatial derivatives are suppressed by very different scales such that $\mathcal{A}_{\rm EFT}$ breaks down before \eqref{eqn:gmax}, could violate~\eqref{eqn:dt_pos2} without sacrificing unitarity/causality/locality. 
The $\gamma^2 = 1$ positivity bound \eqref{eqn:dt_pos1}, on the other hand, is more robust because this constraint could only be evaded if the EFT breaks down already at $\gamma = 1$, i.e. already in the centre-of-mass frame, in which there is no overall motion with respect to the symmetry-breaking background. 

For the particular case of the covariant Galileon discussed in the main text, in practice the vast majority of the constraining power comes from the positivity bounds evaluated at $\gamma^2 < 10$, as illustrated in Figure~\ref{fig-regionPlotMappingComp}. So although in the main text we introduced bound \eqref{eqn:dt_pos2} as a large $\gamma^2$ limit, our subsequent discussion of how this constraint interfaces with observational constraints actually applies to any standard UV completion which produces a EFT able to resolve relative motions up to $\gamma^2 \approx 10$.     

\comment{
\begin{figure}[t!]
\begin{center}
\includegraphics[width=.24\linewidth]{posReg1.pdf}
\includegraphics[width=.24\linewidth]{posReg15.pdf}
\includegraphics[width=.24\linewidth]{posReg110.pdf}
\includegraphics[width=.24\linewidth]{posRegBoth.pdf}
\end{center}
\caption{
Same as figure 3, except without data contours. Coloured regions satisfy stability bounds, while blue regions are the subset excluded by (from left to right): positivity bound (with $\gamma = 1$) only, positivity + `mild velocity 1' ($\gamma = 2$) bound, positivity + `mild velocity 2' ($\gamma = 10$) bound, positivity + `velocity' ($\gamma \to \infty$) bound.
\label{fig-regionPlotMappingComp}}
\end{figure}
}

\begin{figure}[t!]
\begin{center}
\includegraphics[width=.328\linewidth]{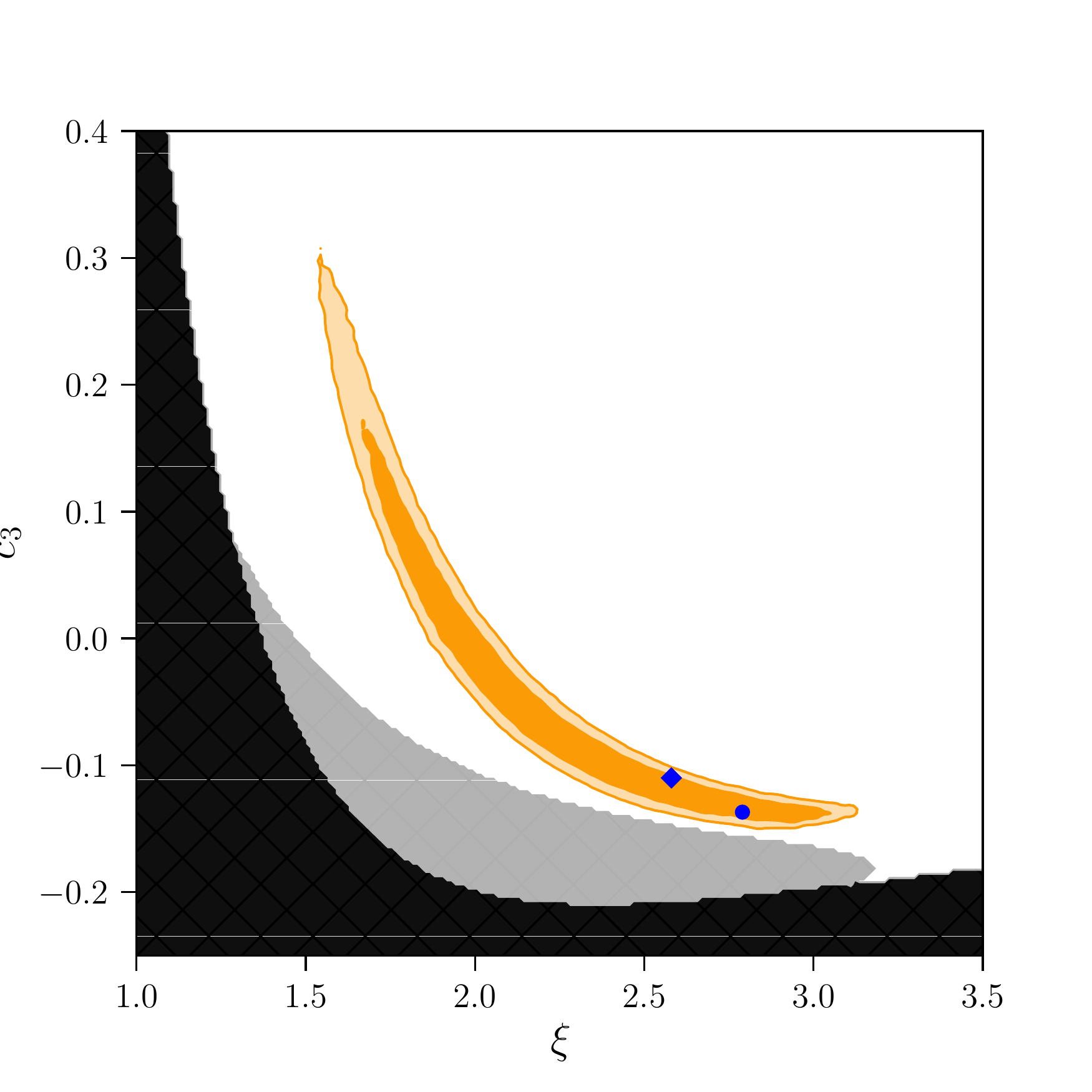}
\includegraphics[width=.328\linewidth]{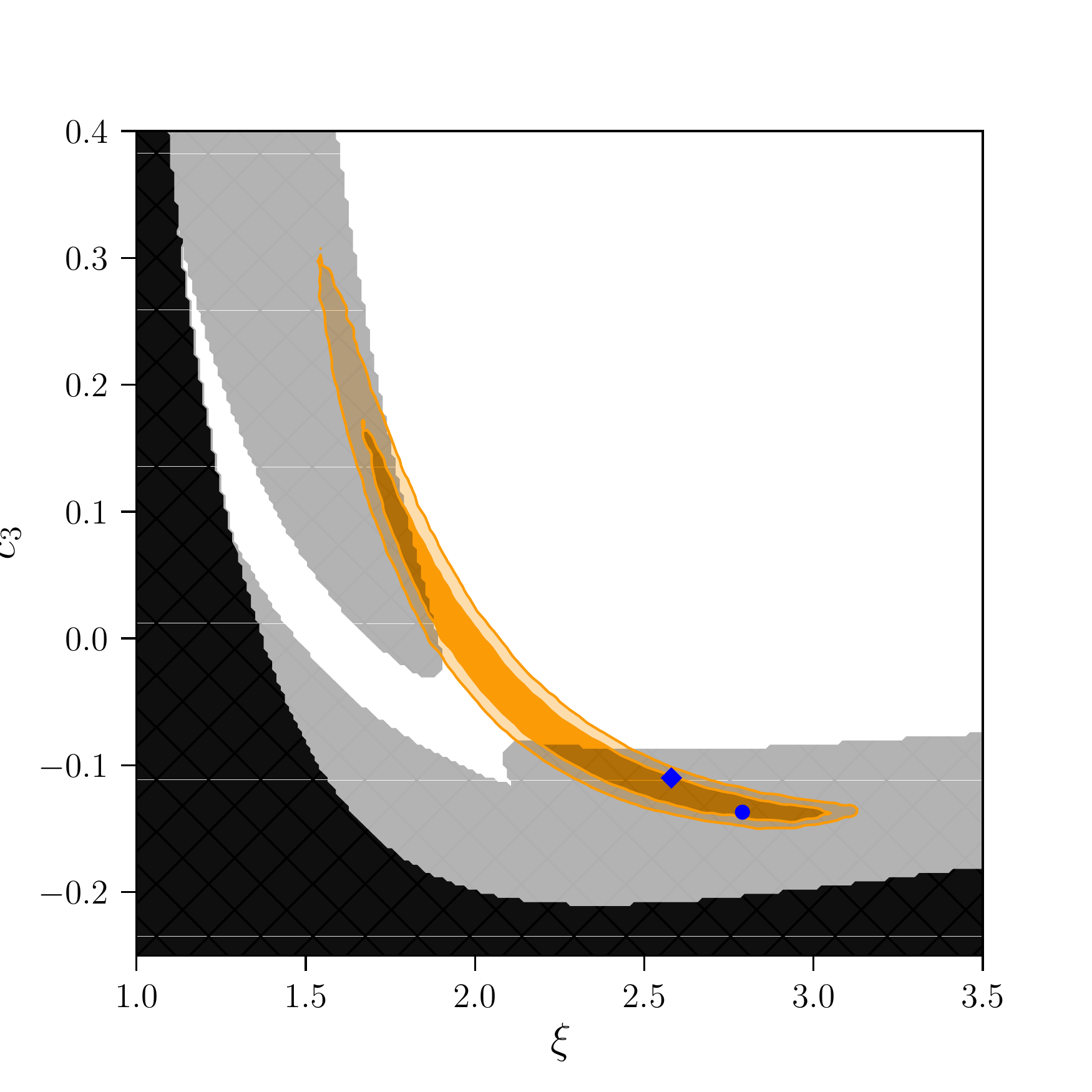}
\includegraphics[width=.328\linewidth]{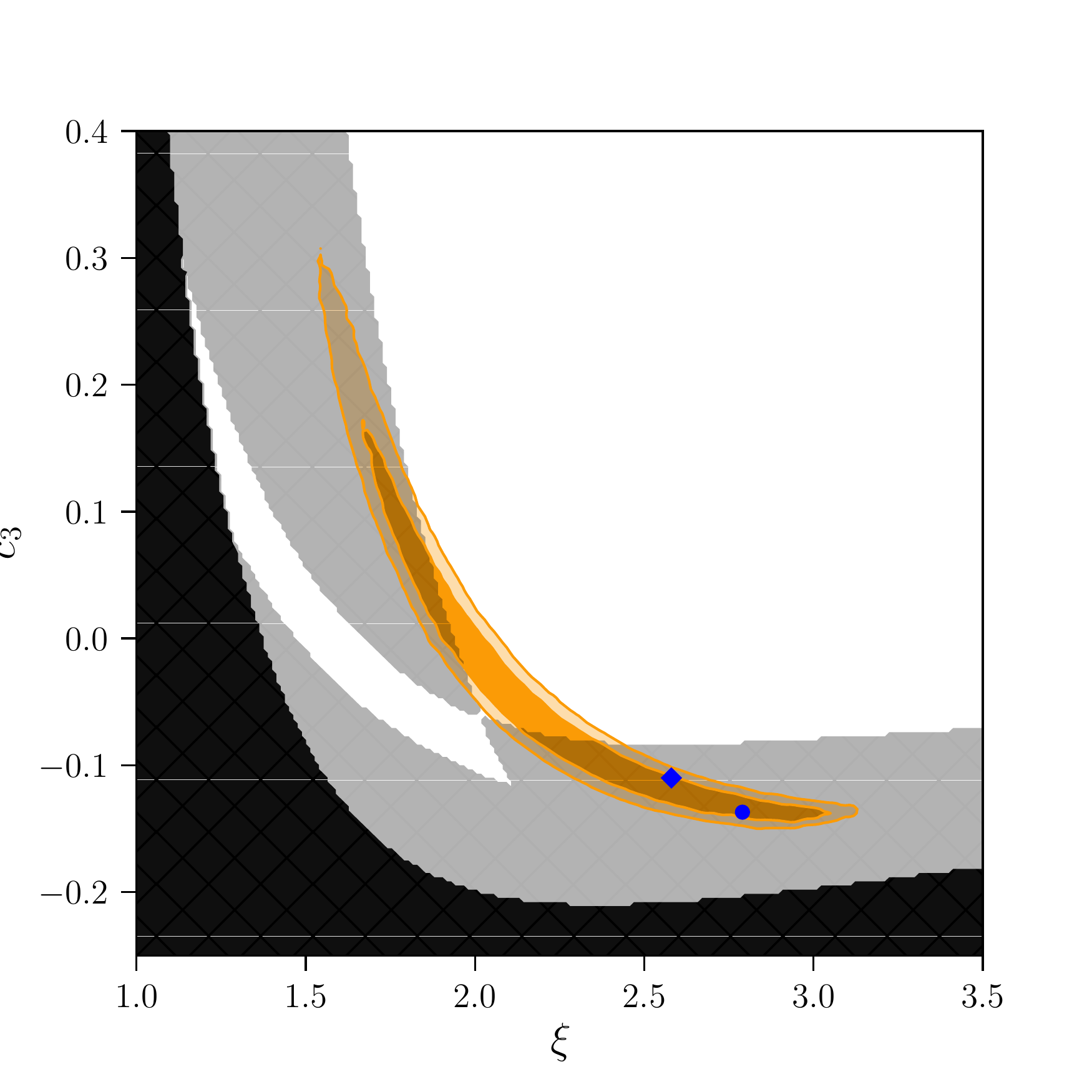}
\end{center}
\caption{
Same as figure 3, but for different values of $\gamma^2$ when evaluating positivity bounds. Black regions fail ghost and/or gradient stability bounds, while grey hashed regions are  excluded by (from left to right): pos. bound 1 \eqref{eqn:dt_pos1} (which sets $\gamma^2 = 1$) only, pos. bound 1 \eqref{eqn:dt_pos1} and the general positivity bound \eqref{eqn:posNLO_csst2} with $\gamma^2 = 10$, pos. bound 1 \eqref{eqn:dt_pos1} and pos. bound 2 \eqref{eqn:dt_pos2} (which effectively sends $\gamma^2 \to \infty$). The observationally relevant positivity priors for the case considered here can therefore already be derived from the small $\gamma$ region, $\gamma^2 \lesssim 10$. 
\label{fig-regionPlotMappingComp}}
\end{figure}

\section{Bounds on a general shift-symmetric amplitude}
\label{app:noGalSymm}

Without a Galileon symmetry, the $2 \to 2$ amplitude will generally depend on more combinations of $\{ s, t, \omega_s, \omega_t, \omega_u\}$ at this order than the two given in~\eqref{eqn:Aexp2}. 
In this section we provide a complete, manifestly crossing symmetric, basis which captures all possible amplitudes, and write down the corresponding positivity bounds at $\mathcal{O} (s^2)$ and $\mathcal{O} (s^2 t)$.

\paragraph{Counting the invariants:}
We begin by simply counting the total number of independent terms which can appear in $\mathcal{A}$ at a given order in derivatives. 
To simplify this counting, we can treat the Mandelstam $u$ as an independent variable (and set $u = -s -t$ at the end of the calculation). Then since we are scattering identical fluctuations, the function $\mathcal{A} (s,t,u, \omega_s, \omega_t, \omega_u)$ must be invariant under the following permutations,
\begin{align}
& \text{Swapping } 1 \leftrightarrow 3: \;\;\;\;  &s \leftrightarrow u \;, \;\;\;&\omega_s \leftrightarrow  \omega_u \\
& \text{Swapping } 2 \leftrightarrow 3: \;\;\;\;  &s \leftrightarrow t \;, \;\;\;&\omega_s \leftrightarrow  \omega_t  \nonumber \\
& \text{Swapping } 1 \leftrightarrow 4: \;\;\;\;  &s \leftrightarrow t \; , \;\;\;&\omega_s \leftrightarrow - \omega_t \; ,
\end{align}
and any combination thereof. We will also focus on interactions that are invariant under time-reversal, so that $\mathcal{A}$ is unchanged by flipping the signs of all energies, $\omega \to - \omega$. 
Altogether, this gives 48 separate permutations under which $\mathcal{A}$ must be invariant. 
Since $\mathcal{A}$ at tree-level is an analytic function of its arguments, we arrive at a simple combinatorics problem: how many ways can one combine powers of the variables $\{s,t,u, \omega_s, \omega_t, \omega_u \}$ to make a term which is invariant under all of the desired permutations? 
Fortunately, the solution to this kind of problem is well-documented. 
We can express the answer in terms of a Hilbert series,
\begin{align}
\mathcal{H} (p, q) &= \sum_{a,b} N_{ab} \, p^a q^b  \, ,
\end{align}
where $N_{ab}$ is the number of invariants that contain $a$ Mandelstam variables (i.e. $a$ powers of $s,t$ or $u$) and $b$ energies (i.e. $b$ powers of $\omega_s, \omega_t$ or $\omega_u$).
Using Molien's formula with our particular set of 48 permutations gives a Hilbert series, 
\begin{align}
 \mathcal{H} (p,q) &= \frac{1 + p q^2 + p q^4 +  p^2 q^2 + p^2 q^4 +  p^3 q^6 }{ (1-p) (1-p^2) (1-p^3) (1-q^2)(1-q^4)(1-q^6)} \label{eqn:Hilbert} \\
 &= 1 + \left( p + q^2 \right) + \left( 2 p^2 + 2 p q^2 + 2 q^4 \right) + ...   \nonumber \, ,
\end{align}
which tells us that there are 2 combinations of mass-dimension 2 (one of the form $s$ and one of the form $\omega^2$, namely $s+t+u$ and $\omega_s^2 + \omega_t^2 + \omega_u^2$), there are 6 combinations of mass-dimension 4 (two of the form $s^2$, two of the form $s \omega^2$, and two of the form $\omega^4$), and so on\footnote{
To account for the constraint $s+t+u =0$, one can simply multiply $\mathcal{H}$ by a factor of $(1-p)$.
}.

\paragraph{An explicit basis:}
In fact, the Hilbert series \eqref{eqn:Hilbert} tells us that \emph{any} polynomial of these six variables which is invariant under these permutations can be written in the following form (a so-called ``Hironaka decomposition'')
\begin{align}
 \mathcal{A} ( s, \, t,  \, u , \, \omega_s, \, \omega_t, \, \omega_u ) =  
\sum_{S_{ij}}^6 \mathcal{A}_{ij} ( P_{ab} ) S_{ij} \; , 
\label{eqn:Hironaka}
\end{align}
where the sum is over the six ``secondary'' generators\footnote{
Note that the secondary generators must be constructed with care, since they require the property that any product $S_{ab} S_{ij}$ can be written in the form \eqref{eqn:Hironaka}.
},
\begin{align}
S_{00} &= 1  \; , &S_{12} &= s \omega_s^2 + t \omega_t^2 + u \omega_u^2  \nonumber  \\
 S_{14} &= s \omega_s^4 + t \omega_t^4 + u \omega_u^4 \;, 
 &S_{22} &= s^2 \omega_s^2 + t^2 \omega_t^2 + u^2 \omega_u^2 \; ,   \\
 S_{24} &= S_{12} S_{12} \; , &S_{36} &= S_{14} S_{22} \; .  \nonumber 
 \end{align}
which may only appear in single powers (this accounts for six terms in the numerator of \eqref{eqn:Hilbert}), with coefficients that depend on six ``primary'' generators,
\begin{align}
 P_{10} &= s + t + u \; , &P_{02} &= \omega_s^2 + \omega_t^2 + \omega_u^2 \; ,  \nonumber \\
 P_{20} &= s^2 + t^2 + u^2 \;, &P_{04} &= \omega_s^4 + \omega_t^4 + \omega_u^4 \; ,  \\ 
 P_{30} &= s^3 + t^3 + u^3 \; , &P_{06} &= \omega_s^6 + \omega_t^6 + \omega_u^6 \; ,  \nonumber 
 \end{align}
which may appear to arbitrary powers, $(P_{ab})^N$ for any $N$ (this accounts for the six factors in the denominator of \eqref{eqn:Hilbert}). 
Finally, imposing $s+t+u = 0$ corresponds to setting $P_{10} = 0$ and removes one of the primary generators. 

Explicitly, this reveals that the most general tree-level amplitude (invariant under spacetime translations and spatial rotations, but not boosts) is described up to sixth order in derivatives by just 14 Wilson coefficients,  
\begin{align}
& \mathcal{A} (s, \, t , \, \omega_s \, , \omega_t \, , \omega_u )   \label{eqn:Ageneral} \\
 &= c_{00} + c_{02} P_{02}    
 + c_{20} P_{20} + c_{12} S_{12} + c_{04} P_{04} + c_{04}' P_{02}^2 
 \nonumber \\
 &+ c_{30} P_{30} + c_{22} S_{22} + c_{22}' P_{20} P_{02}
 + c_{14} S_{14} + c_{14}' P_{02} S_{12}
 + c_{06} P_{06} + c_{06}' P_{04} P_{02} + c_{06}'' P_{02}^3 .  \nonumber 
\end{align}
Using this general template, we are now going to impose an approximate shift-symmetry to wittle the number of free coefficients down a little further, and then apply positivity bounds.

\paragraph{At leading order:}
An exact shift symmetry would require that the amplitude exhibits certain soft behaviour, in particular $\mathcal{A} \to 0$ when any single $p_a \to 0$ (equivalently, the Lagrangian is built from interactions in which each field is covered by at least one derivative). 
Even when this symmetry is broken by a small mass, this soft behaviour in the massless limit can be used to infer a hierarchy between the coefficients in \eqref{eqn:Ageneral}.
For instance, the $c_{00}$ and $c_{02}$ terms in \eqref{eqn:Ageneral} are suppressed (since they correspond to interactions with fewer than one derivative per field), and so the leading order amplitude begins at mass-dimension 4. Furthermore, only one combination of $c_{04}$ and $c_{04}'$ is unsuppressed---namely the term $\omega_1 \omega_2 \omega_3 \omega_4$ (in which each field has a single derivative). This means that a general shift-symmetric amplitude, at leading order in derivatives, is described by just three Wilson coefficients,
\begin{align}
 \mathcal{A}_{\rm LO} (s, \, t  , \, \omega_s , \, \omega_t , \, \omega_u ) 
 = \frac{1}{2} c_{ss} \left( s^2 + t^2 + u^2 \right) + \frac{1}{4} d_{s \omega \omega} \left( s \omega_s^2 + t \omega_t^2 + u \omega_u^2  \right)    + d_{\omega \omega \omega \omega} \omega_1 \omega_2 \omega_3 \omega_4
 \label{eqn:ALO}
\end{align}
where we have redefined the coefficients $\{ c_{20} , c_{12}, c_{04} \}$ as ${ c_{ss}, d_{s\omega\omega}, d_{\omega \omega \omega \omega} }$ to match \eqref{eqn:ABoostlessFwd}. 
Positivity of the UV completion requires that these coefficients obey the bound \eqref{eqn:posLO_css3} given in the main text.

\paragraph{At next-to-leading order:}
At next-to-leading order (i.e. at sixth order in derivatives), there are up to 8 new coefficients, but only 6 of these are compatible with an approximate shift-symmetry. Separating out the Galileon-invariant terms (and renaming their coefficients to $c_{sst}$ and $d_{s t \omega \omega}$ to match \eqref{eqn:Aexp2}), we have that the most general tree-level EFT amplitude at sixth order in derivatives (and constrained by spacetime translations, spatial rotations, and an approximate shift symmetry) is given by,   
\begin{align}
& \mathcal{A}_{\rm NLO} (s, \, t , \, \omega_s \, , \omega_t \, , \omega_u )   \label{eqn:ANLO} \\
&= 
 - \frac{1}{3} c_{sst} \left( s^3+ t^3+u^3 \right) 
 - \frac{1}{4} d_{st\omega\omega} \left( s^2 ( \omega_1 \omega_2 + \omega_3 \omega_4) + t^2 ( \omega_1 \omega_3 + \omega_2 \omega_4) + u^2 (\omega_1 \omega_4 + \omega_2 \omega_3 ) \right) \nonumber \\
&+ c_{22} \left( s^2 \omega_s^2 + t^2 \omega_t^2 + u^2 \omega_u^2 \right) 
 + c_{14} \left( s \omega_s^4 + t \omega_t^4 + u \omega_u^4  \right) 
 + c_{14}' \left( s \omega_s^2 + t \omega_t^2 + u \omega_u^2  \right) \left( \omega_s^2 + \omega_t^2 + \omega_u^2   \right) \nonumber \\
&+ c_{06} \; \omega_1 \omega_2 \omega_3 \omega_4 \left( \omega_s^2 + \omega_t^2 + \omega_u^2  \right) .  \nonumber 
\end{align}
It is easy to see that the $c_{14}, c_{14}'$ and $c_{06}$ terms could never arise from a Galileon invariant theory, since the only interactions with that many derivatives require an antisymmetric derivative structure to be invariant, which can produce at most two time derivatives. The $c_{22}$ term cannot be produced for essentially the same reason: the antisymmetric structure requires that the amplitude vanishes in the forward limit.

Accounting for the four additional Wilson coefficients in \eqref{eqn:ANLO}, as well as the three Wilson coefficients in \eqref{eqn:ALO}, which are not Galileon invariant but which can generally arise in any shift-symmetric scalar field theory, the positivity bound \eqref{eqn:posNLO} becomes an inequality that depends on $\gamma$, $M$ and $\delta$.
Taking $\gamma^2 M \ll \delta$ gives a bound,
\begin{align}
 c_{sst}  + \gamma^2 \left( \frac{1}{4} d_{st\omega \omega} - c_{22} \right)  -  \gamma^4 \left( \frac{3}{8} c_{14} + \frac{1}{4} c_{14}' \right) > - \frac{3}{2 s_b} \left(  c_{ss} + \frac{1}{4} d_{s\omega \omega} \gamma^2 + \frac{1}{16} d_{\omega \omega \omega \omega} \gamma^4  \right) \; , 
\end{align}
which reduces to the bound \eqref{eqn:posNLO_csst2} studied in the main text once restricted to only the Galileon invariant terms.

\FloatBarrier

\bibliographystyle{JHEP}
\bibliography{pos_CovGal}
\end{document}